\documentclass[twocolumn,aps,pre,longbibliography]{revtex4-1}

\usepackage{epsf}
\usepackage{subfigure}
\usepackage[colorlinks, linkcolor =blue]{hyperref} 
\usepackage{amsmath}
\usepackage{amssymb}
\usepackage{graphicx}
\usepackage{epstopdf}
\usepackage{color}
\usepackage{mathrsfs}

\newcommand{\be}{\begin{equation}}
\newcommand{\ee}{\end{equation}}
\newcommand{\bea}{\begin{eqnarray}}
\newcommand{\eea}{\end{eqnarray}}

\begin{document}


\title{Microreversibility and quantum transport in Aharonov-Bohm rings}

\author{M. Barbier and P. Gaspard}
\affiliation{Center for Nonlinear Phenomena and Complex Systems,\\
Universit\'e Libre de Bruxelles (ULB), Code Postal 231, Campus Plaine, 
B-1050 Brussels, Belgium}


\begin{abstract}
The consequences of microreversibility for the linear and nonlinear transport properties of systems subjected to external magnetic fields are systematically investigated in Aharonov-Bohm rings connected to two, three, and four terminals.  Within the independent electron approximation, the cumulant generating function, which fully specifies the statistics of the nonequilibrium currents, is expressed in terms of the scattering matrix of these circuits.  The time-reversal symmetry relations up to the third responses of the currents and the fourth cumulants are analytically investigated and numerically tested as a function of the magnetic flux. The validity of such relations is thus firmly confirmed in this class of open quantum systems.
\end{abstract}

\noindent 
\vskip 0.5 cm

\maketitle


\section{Introduction}
\label{intro_sec}

Nonlinear transport properties manifest themselves if the thermodynamic forces driving the system away from equilibrium are applied over distances shorter than the mean free path of the particles. This is for instance the case in mesoscopic electronic circuits where electrons have ballistic motion, so that transport properties strongly deviate from Ohm's law of proportionality between currents and voltage differences~\cite{Imry,Naz}.

Moreover, multi-terminal circuits allow the coupling between several electric currents. In the vicinity of equilibrium, this coupling is described by the conductance coefficients, which are proportionality factors between the electric currents and the applied voltage differences between the terminals \cite{Dat}. As a consequence of microreversibility, i.e., the symmetry of the microscopic dynamics under the time-reversal transformation, these coefficients obey the well-known Onsager-Casimir reciprocal relations and the fluctuation-dissipation theorem~\cite{Ons31a,Ons31b,Cas45,CW51}. Their domain of validity is restricted to the linear regime close to equilibrium. However, large voltage differences can be implemented into mesoscopic electronic circuits, hence making the latter typically operate in nonlinear regimes far away from equilibrium.  In this context, the question arises about the consequences of microreversibility on the nonlinear transport properties, beyond Ohm's law.

Remarkably, great advances have been achieved about this issue \cite{BC59,BK77,S92,S94}, especially, with the advent of the so-called {\it fluctuation relations}, which are time-reversal symmetry relations among the probability distributions of opposite fluctuations in transport properties \cite{ECM93,ES94,GC95,Kur98,LS99,Cro99,J00,Kur00,Gas04,TN05,S12}.  Furthermore, these relations have been extended to the full counting statistics of all the currents flowing across an open system \cite{AG06,AG07JSP,SU08,AGM09,EHM09,CHT11,HPPG11,Gas13NJP,Gas15}.  Such multivariate fluctuation relations allow us to deduce not only the Onsager-Casimir reciprocal relations and the fluctuation-dissipation theorem, but also their generalizations beyond the linear regime for currents in the absence or the presence of an external magnetic field \cite{AG04,AG07JSM,SU08,AGM09,Gas13NJP,Gas13KJJ,LLS12,WF15}.  Recently, these generalized time-reversal symmetry relations have been analyzed at arbitrarily high orders in the thermodynamic forces, also called affinities, that generate the currents \cite{BG18,BG19,BG20}.

In this paper, our purpose is to investigate systematically these generalized relations between the first, second, third, and fourth cumulants and their responses to nonequilibrium constraints in the specific case of Aharonov-Bohm rings connected to several terminals. The generalized relations have already been obtained and explicitly written down up to third cumulants for two-terminal circuits with a single current driven by a single affinity, as well as for multi-terminal circuits where two affinities are varied and the cumulants of the two corresponding currents are considered~\cite{SU08}. These relations have been theoretically studied in a two-terminal Aharonov-Bohm interferometer up to third cumulants \cite{US09}. Moreover, coherent quantum transport in two-terminal Aharonov-Bohm rings has also been experimentally probed up to the second response of the current and the first response of the diffusivity \cite{NYH10,NYH11}.  However, the theoretical study of the generalized relations at arbitrary orders shows an alternance between even and odd orders \cite{BG18,BG19,BG20}. In this regard, it is important to investigate together the nonlinear properties associated with the second and third responses of the currents, or the third and fourth cumulants, and in circuits with more than two or three terminals, which is here our aim.

The plan of the paper is the following.  In Sec.~\ref{Quantum}, the cumulant generating function is deduced for open systems of non-interacting fermions in terms of the scattering matrix describing quantum transport of single fermions in the presence of an external magnetic field. Moreover, the multivariate fluctuation relation satisfied by this generating function as a consequence of microreversibility is presented.  In Sec.~\ref{transport_prop_sec}, the consequences of the multivariate fluctuation relation on the transport properties of the system are then systematically analyzed up to the third responses of the currents and the fourth cumulants.  These results are applied to Aharonov-Bohm rings in Sec.~\ref{AB_sec}.  Conclusions are finally drawn in Sec.~\ref{Conclusion}.


\section{Quantum transport of independent electrons in multi-terminal circuits}
\label{Quantum}

In this section, we recall some of the features of quantum transport that prove relevant for our subsequent analysis. We begin in Subsec.~\ref{multi_term_subsec} with a brief review of the class of systems known as multi-terminal circuits. The assumption of independent electrons allows us to describe their dynamics within such a system by means of a mere one-body Hamiltonian operator. A multi-terminal circuit is a prototypical example of a nonequilibrium system, in which currents of energy or particles occur. The latter having a statistical origin, they are intrinsically random variables that require a stochastic description, as we discuss in Subsec.~\ref{FCS_subsec}. We introduce in particular the generating function of the statistical cumulants, which is seen in Subsec.~\ref{FR_subsec} to satisfy a multivariate fluctuation relation. 


\subsection{Multi-terminal circuits}
\label{multi_term_subsec}

\begin{figure}[t]
\begin{center}
\includegraphics[scale=0.25]{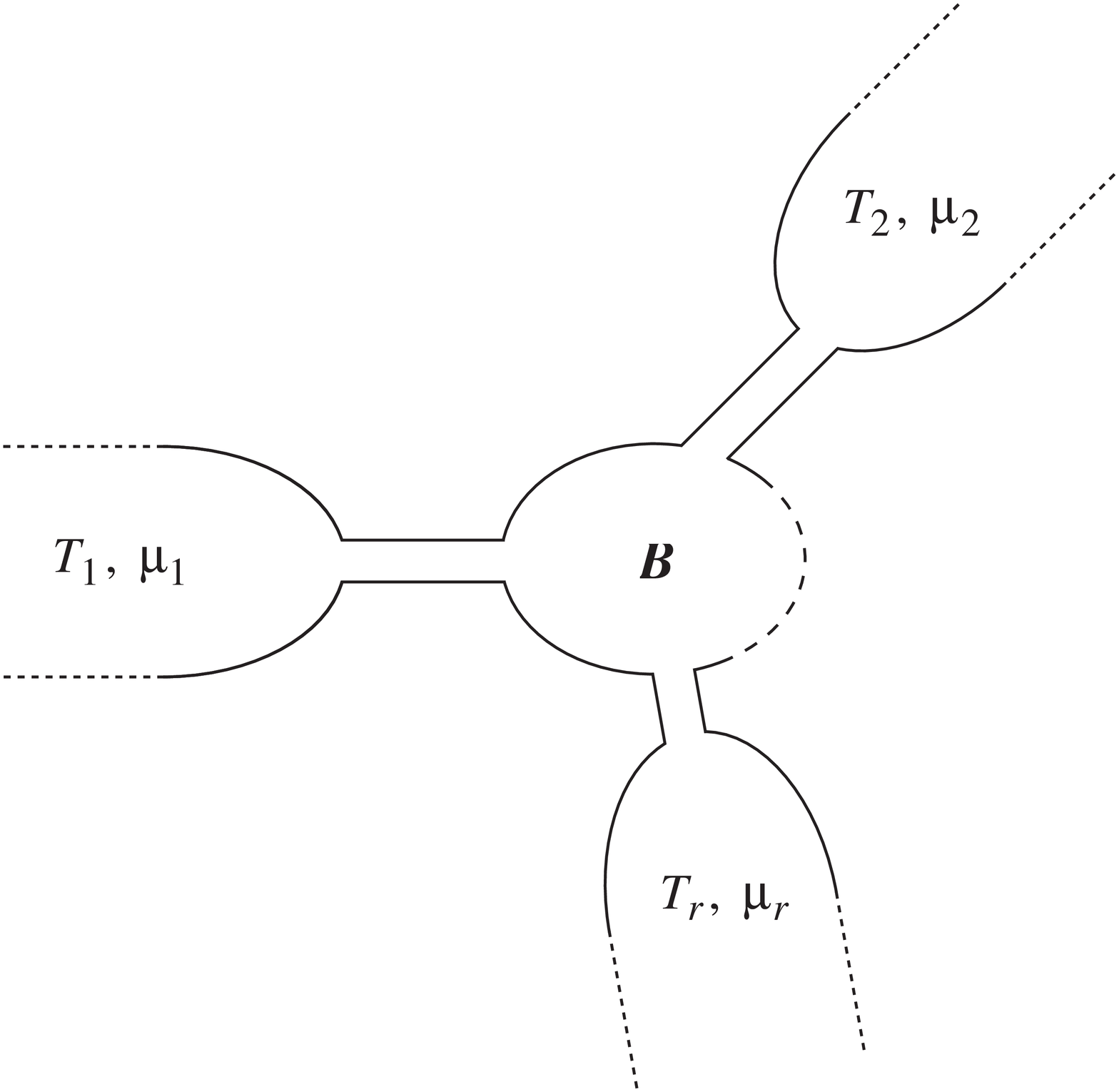}
\end{center}
\caption{Schematic representation of a multi-terminal circuit, where $r$ reservoirs of electrons, $r \geqslant 2$, are coupled through $r$ terminals. Each terminal connects a reservoir to a central region. We assume that the external magnetic field $\boldsymbol{B}$ only acts within the central region, and hence not within the reservoirs and the terminals.}
\label{fig1}
\end{figure}

We consider a system formed by independent electrons that are subjected to an external static magnetic field~$\boldsymbol{B}$. The many-body Hamiltonian operator $\hat H(\boldsymbol{B})$ that governs the dynamics of the system can be decomposed into the position representation as (see e.g. Ref.~\cite{Bruus})
\be
\hat H(\boldsymbol{B}) = \sum_{\sigma=\pm} \int  d{\bf r} \, \hat\psi_{\sigma}^{\dagger}({\bf r}) \, \hat h(\boldsymbol{B}) \, \hat\psi_{\sigma}({\bf r})
\label{H_space}
\ee
in terms of the anticommuting field operators $\hat\psi_{\sigma}({\bf r})$, with the vector $\bf r$ representing the position in three-dimensional space and $\sigma$ labelling the two possible values of the spin of an electron. The expression~\eqref{H_space} of the total Hamiltonian only involves the one-body Hamiltonian operator $\hat h(\boldsymbol{B})$ as a result of the independence of the electrons. We have
\be
\hat h(\boldsymbol{B}) = \frac{1}{2m} \left[ -i \hbar \pmb{\nabla} +  e \boldsymbol{\mathcal{A}} \left( {\bf r} ; \boldsymbol{B} \right) \right]^2 + u({\bf r}) \, ,
\label{general_one_body_Ham}
\ee
with $m$ and $-e$ the mass and the electric charge, respectively, of an electron, $\boldsymbol{\mathcal{A}} \left( {\bf r} ; \boldsymbol{B} \right)$ the vector potential associated with the magnetic field $\boldsymbol{B}$, i.e., such that $\boldsymbol{B} = \pmb{\nabla} \times \boldsymbol{\mathcal{A}}$, and $u({\bf r})$ an additional potential that constrains the dynamics of the electron. We consider the case where the total system forms a so-called multi-terminal circuit, as illustrated in Fig.~\ref{fig1}.

A multi-terminal circuit consists in $r \geqslant 2$ reservoirs of electrons, which are supposed to be prepared in their state of thermodynamic equilibrium at the temperatures $T_j$ and the chemical potentials $\mu_j$ with $j = 1 , \ldots , r$. We consider ideal (infinitely large) reservoirs, so that their thermodynamic properties do not change in the course of time. These $r$ reservoirs are coupled through $r$ terminals to a central region where the electrons can be scattered into the different terminals. This scattering region can represent various physical devices, typical examples in quantum transport being a quantum point contact, a quantum dot, or an Aharonov-Bohm ring. The external magnetic field $\boldsymbol{B}$ is taken to act only within the scattering region. In particular, we thus have $\boldsymbol{B} = \boldsymbol{0}$ both in the reservoirs and the terminals. For brevity, we gather in the sequel the reservoirs with the terminals, and will generically refer to them as terminals. Therefore, the one-body Hamiltonian operator~\eqref{general_one_body_Ham} can be written in the form
\begin{widetext}
\be
\hat h(\boldsymbol{B}) = \left\{\begin{array}{ll}
\hat h_{\mathrm{scat}}(\boldsymbol{B}) = \frac{1}{2m} \left[ -i \hbar \pmb{\nabla} + e \boldsymbol{\mathcal{A}} \left( {\bf r} ; \boldsymbol{B} \right) \right]^2 + u({\bf r}) \, , & \qquad \text{in the scattering region,} \\[0.5cm]
\hat h_{j} = -\frac{\hbar^2}{2 m} \pmb{\nabla}^2 + u_{j} ({\bf r}) \, , &\qquad \text{in the terminal $j$.}
\end{array} \right.
\label{one_body_Ham_split}
\ee
\end{widetext}

The terminals act as waveguides for the transport of electrons. We assume that the motion of an electron in each terminal is unbounded in the longitudinal direction $x$ and confined by a potential $u_{j} ({\bf r}) = u_{\perp} (y,z)$ in the transverse directions $y$ and $z$. The electron is then taken to behave as a free particle in the longitudinal direction $x$ along the terminal. We emphasize that the directions $x,y$ and $z$ are specific to each terminal $j$ ($j=1, \ldots , r$).  In particular, all the terminals have their longitudinal direction oriented along a semi-infinite axis that is denoted by $x$, with $x \in \left[ 0, \infty \right)$. The Hamiltonian operator $\hat h_{j}$ in Eq.~\eqref{one_body_Ham_split} hence reduces to
\be
\hat h_{j} = -\frac{\hbar^2}{2 m} \pmb{\nabla}^2 + u_{\perp} (y,z) \, , \qquad \text{in each terminal $j$.}
\label{Ham_term}
\ee
Since the Hamiltonian operator~\eqref{Ham_term} is separable in the longitudinal and transverse directions $x$ and $(y,z)$, respectively, it admits eigenstates of the form (see e.g. Ref.~\cite{Naz})
\be
\hat h_{j} \, \psi_{j} (x) \, \phi_{n_y , n_z} (y,z) = \left( \varepsilon_k + E_{n_y , n_z} \right) \psi_{j} (x)\, \phi_{n_y , n_z} (y,z) \, ,
\label{eigenstates_term}
\ee
with the free-particle eigenenergy $\varepsilon_k = \hbar^2 k^2 / 2m$. The longitudinal eigenstates $\psi_{j} (x)$ are mere linear combinations of plane waves, i.e.,
\be
\psi_{j} (x) = a_j \, \mathrm{e}^{-ikx} + b_j \, \mathrm{e}^{ikx} \, ,
\label{free_eigenstates}
\ee
while the transverse eigenstates $\phi_{n_y , n_z} (y,z)$, as well as the corresponding energies $E_{n_y , n_z}$, depend on the specific form of the confining potential $u_{\perp} (y,z)$. An important point however is that, while the wavenumber $k$ is continuous, the two indices $n_y$ and $n_z$ are typically discrete.

The transverse motion of an electron in a terminal is thus quantized into modes, a mode corresponding to a specific pair of indices $\left\{ n_y,n_z \right\}$. In the context of quantum transport, a mode is often referred to as a transport channel. The minimum energy of the channel $\left\{ n_y,n_z \right\}$ is given by $E_{n_y , n_z}$, since $\varepsilon_k \geqslant 0$. An electron of energy~$E$ can only be in a superposition of the eigenstates~\eqref{eigenstates_term} that satisfy $\varepsilon_k + E_{n_y , n_z} \leqslant E$. In particular, any channel $\left\{ n_y,n_z \right\}$ for which $E_{n_y , n_z} > E$ is said to be closed. Accordingly, the transport only concerns the accessible, so-called open channels. Here and in the sequel, we consider the case where only a single channel is open. This description of the transport process is of course only valid in the terminals, i.e., far from the scattering region. The effect of the latter is to scatter an electron between the different terminals.

The scattering region can be described by means of the scattering unitary operator $\hat S(\varepsilon;\boldsymbol{B})$, which depends on the magnetic field $\boldsymbol{B}$. Most generically, the scattering operator relates, in a one-to-one way, the asymptotic state after the scattering event to the asymptotic state before the scattering event \cite{Tay}. As we stated above, the system considered in this work is a $r$-terminal circuit with a separable potential and a single open channel in each of the terminals. In this context, the scattering operator $\hat S(\varepsilon;\boldsymbol{B})$ is a $r \times r$ matrix that expresses the coefficients $b_j$ of the longitudinal eigenstates~\eqref{free_eigenstates} in terms of the coefficients $a_j$. In other words, it relates the plane waves propagating away from the scattering region [namely, $\exp(ikx)$] to the plane waves incoming the scattering region [i.e., $\exp(-ikx)$]. An important feature of this scattering matrix is that it can be used to determine the statistical properties of the currents of energy and electrons that take place within the circuit, as we now discuss.


\subsection{Full counting statistics}\label{FCS_subsec}

The differences of temperatures $T_j$ and chemical potentials $\mu_j$ with $j = 1, \ldots , r$ in the reservoirs make the $r$-terminal circuit here considered to operate out of equilibrium. One of the reservoirs is taken as the reference reservoir, for instance the $r^{\rm th}$ reservoir. The nonequilibrium state is characterized by the occurrence of currents of energy and electrons between the reservoirs.  These currents are driven by the thermodynamic forces defined by
\bea
A_{jE} &=& \frac{1}{k_{\rm B}T_r} - \frac{1}{k_{\rm B}T_j} \, , \label{A_E}\\
A_{jN} &=& \frac{\mu_j}{k_{\rm B}T_j} -\frac{\mu_r}{k_{\rm B}T_r}  \label{A_N} \, ,
\eea
with $j=1,2,...,r-1$ and $k_{\rm B}$ the Boltzmann constant. These thermodynamic forces are commonly referred to as affinities in the literature \cite{DeDon,Prig,deGr,Cal}, a terminology that we use throughout this paper.  The thermal and chemical affinities~\eqref{A_E} and~\eqref{A_N}, respectively, can be gathered into the vector
\be
\boldsymbol{A} = \left( A_{1E} , \ldots , A_{r-1 , E} , A_{1N} , \ldots , A_{r-1 , N} \right)
\label{A_def}
\ee
with $2(r-1)$ components giving all the possible control parameters driving the currents across the circuit. These affinities are vanishing at equilibrium where the temperature and the chemical potential are uniform in the whole system (i.e., $T_j=T_r$ and $\mu_j=\mu_r$ for $j=1,2,...,r-1$).

At the microscopic level of description, the currents have a statistical origin and are thus random variables. These currents can be evaluated through the widely-used two-point measurement scheme \cite{Kur00,AGM09,EHM09}. The latter assumes that the reservoirs are coupled during some time interval $[0,t]$ with $t>0$. Measurements of energy and electron number are performed in each reservoir before the initial time $\tau=0$ and after the final time $\tau=t$. The amounts $\Delta E_j$ and $\Delta N_j$ of energy and electrons exchanged between the $j^{\rm th}$~reservoir with $j=1,...,r-1$ and the reference reservoir $j=r$ during the time interval $[0,t]$ as a result of the coupling are then obtained by taking the differences of the corresponding eigenvalues obtained in the two measurements.

All the $2(r-1)$ quantities $\{\Delta E_j\}_{j=1}^{r-1}$ and $\{\Delta N_j\}_{j=1}^{r-1}$ are collectively denoted by
\begin{multline}
\Delta\boldsymbol{X} \equiv (\Delta\boldsymbol{E},\Delta\boldsymbol{N}) \\
= (\Delta E_1 , \ldots , \Delta E_{r-1} , \Delta N_1 , \ldots , \Delta N_{r-1}) \, .
\label{proba_dis_def}
\end{multline}
These random variables are described by the probability distribution $p_t(\Delta\boldsymbol{X},\boldsymbol{A};\boldsymbol{B})$, which embeds all the statistical properties of the currents of energy and electrons that take place in the nonequilibrium system specified by the affinities~\eqref{A_def}. For this reason, we say that it characterizes the full counting statistics. The latter can be alternatively described by the generating function $G_t$ of the statistical moments \cite{vanK}, related to the probability distribution through the Laplace transform
\be
G_t \left( \pmb{\lambda}, \boldsymbol{A} ; \boldsymbol{B} \right) \equiv \int p_t(\Delta\boldsymbol{X},\boldsymbol{A};\boldsymbol{B})
\, \exp(-\boldsymbol{\lambda}\cdot\Delta\boldsymbol{X}) \, d\Delta\boldsymbol{X} \, ,
\label{mom_GF_def}
\ee
which is a function of the so-called counting parameters (or counting fields) 
\be
\boldsymbol{\lambda} = \left( \lambda_{1E} , \ldots , \lambda_{r-1 , E} , \lambda_{1N} , \ldots , \lambda_{r-1 , N} \right)
\label{lambda_def}
\ee
associated with the energy differences $\Delta\boldsymbol{E}$ and  electron number differences $\Delta\boldsymbol{N}$. We then consider the long-time limit $t \to \infty$ and construct the function $Q$ defined by
\be
Q(\boldsymbol{\lambda},\boldsymbol{A};\boldsymbol{B})  \equiv - \lim_{t \to \infty} \frac{1}{t} \, \mathrm{ln} \, G_t \left(\boldsymbol{\lambda},\boldsymbol{A} ; \boldsymbol{B} \right) \, ,
\label{long_time_lim}
\ee
which is hence nothing but the generating function of the statistical cumulants. 

For systems of non-interacting fermions, using methods developed in Refs.~\cite{Kli,ABGK08}, it can be shown \cite{Gas13KJJ,Gas13NJP,Gas15} that the function defined by Eq.~\eqref{long_time_lim} is given by
\begin{multline}
Q(\boldsymbol{\lambda},\boldsymbol{A};\boldsymbol{B}) = - g_s \int \frac{d\varepsilon}{2\pi\hbar} \, \ln\det\left\{1 \vphantom{\hat f(\varepsilon) \left[\hat S^{\dagger} (\varepsilon;\boldsymbol{B}) \, {\rm e}^{\varepsilon\, \hat\lambda_E + \hat \lambda_N}\, \hat S(\varepsilon;\boldsymbol{B}) \, {\rm e}^{-\varepsilon\, \hat\lambda_E + \hat \lambda_N} - 1 \right]} \right. \\[0.25cm]
\left. + \hat f(\varepsilon) \left[\hat S^{\dagger} (\varepsilon;\boldsymbol{B}) \, {\rm e}^{\varepsilon\, \hat\lambda_E + \hat \lambda_N}\, \hat S(\varepsilon;\boldsymbol{B}) \, {\rm e}^{-\varepsilon\, \hat\lambda_E + \hat \lambda_N} - 1 \right]\right\} \, ,
\label{Q_expr}
\end{multline}
where $g_s$ denotes the spin multiplicity of the particles involved in the nonequilibrium currents, i.e., $g_s=2$ here for electrons \cite{Naz}, and $\hat S^{\dagger} (\varepsilon;\boldsymbol{B})$ is the adjoint of the scattering matrix $\hat S (\varepsilon;\boldsymbol{B})$. In Eq.~\eqref{Q_expr}, the quantity $\hat f(\varepsilon)$ is the $r \times r$ diagonal matrix
\be
\hat f(\varepsilon) = 
\Big( f_j(\varepsilon)\, \delta_{jj'} \Big)_{1 \leqslant j,j' \leqslant r} \, ,
\label{Fermi_Dirac_matrix}
\ee
with $\delta_{jj'}$ the Kronecker delta, and whose elements are the Fermi-Dirac distributions of the reservoirs, namely
\be
f_j(\varepsilon) = \frac{1}{\exp \left[ \left( \varepsilon-\mu_j \right) /(k_{\rm B}T_j)\right] + 1} \, .
\label{Fermi_Dirac_dis}
\ee
Moreover, ${\rm exp} (\varepsilon\, \hat\lambda_E + \hat \lambda_N)$ denotes the diagonal matrix
\be
{\rm e}^{\varepsilon\, \hat\lambda_E + \hat \lambda_N} = 
\Big( {\rm e}^{\varepsilon\, \lambda_{jE} + \lambda_{jN}}\, \delta_{jj'} \Big)_{1 \leqslant j,j' \leqslant r}
\label{exp_matrix_def}
\ee
with $\lambda_{rE}=0$ and $\lambda_{rN}=0$ for the reference reservoir. The symbols $\hat \lambda_E = \left( \lambda_{jE} \, \delta_{jj'} \right)_{1 \leqslant j,j' \leqslant r}$ and $\hat \lambda_N = \left( \lambda_{jN}  \, \delta_{jj'} \right)_{1 \leqslant j,j' \leqslant r}$ are thus the $r \times r$ matrices containing on their diagonal, respectively, the counting parameters $\lambda_{jE}$ for energy exchanges and $\lambda_{jN}$ for electron transfers. The expression~\eqref{Q_expr} of the cumulant generating function is equivalent to the Levitov-Lesovik formula \cite{LL93}, as can be checked for two-terminal circuits \cite{Gas13KJJ}.

It is worth noting that the dependence of the function~$Q$ on the variables $\boldsymbol{\lambda}$ and $\boldsymbol{A}$ defined by Eqs.~\eqref{lambda_def} and~\eqref{A_def}, respectively, can be readily seen from its expression~\eqref{Q_expr}.  In particular, the $\boldsymbol{\lambda}$-dependence is rooted in the fact that the $r \times r$ determinant in Eq.~\eqref{Q_expr} is invariant under the transformations $\exp(\varepsilon\, \hat\lambda_E + \hat \lambda_N) \to \exp[ \varepsilon\, \hat\lambda_E + \hat \lambda_N + \chi(\varepsilon)\, \hat 1]$, where $\chi(\varepsilon)$ is an arbitrary function of the energy $\varepsilon$ and $\hat 1$ denotes the $r \times r$ identity matrix. Each particular function $\chi(\varepsilon)$ allows the definition of a particular set of $2(r-1)$ independent currents of energy and electrons by taking another reference reservoir than $j=r$, as is for instance done in Ref.~\cite{AGM09}. The dependence of the function $Q$ on the affinities $\boldsymbol{A}$ then rises from the Fermi-Dirac matrix $\hat f(\varepsilon)$ in Eq.~\eqref{Q_expr}.  In addition to $\boldsymbol{A}$, the function~$Q$ also depends on the temperature $T_r$ and the chemical potential $\mu_r$ of the reference reservoir $j=r$.


\subsection{The time-reversal symmetry and the multivariate fluctuation relation}\label{FR_subsec}

In the presence of an external magnetic field $\boldsymbol{B}$ and for spinless particles (or for each spin component in systems without spin-orbit interaction), the time-reversal symmetry implies that the scattering matrix satisfies \cite{Dat,LS11}
\be
\hat S^{\rm T}(\varepsilon;\boldsymbol{B}) = \hat S(\varepsilon;-\boldsymbol{B}) \, ,
\label{S-symm}
\ee
the superscript ${\rm T}$ denoting the transpose. Consequently, the cumulant generating function~\eqref{Q_expr} obeys the following time-reversal symmetry:
\be
Q(\boldsymbol{\lambda},\boldsymbol{A};\boldsymbol{B}) = Q(\boldsymbol{A}-\boldsymbol{\lambda},\boldsymbol{A};-\boldsymbol{B}) \, ,
\label{sym-Q-magn}
\ee
which is the expression of the multivariate fluctuation relation \cite{Gas13NJP}. Its consequences on the linear and nonlinear transport properties of the system are given below in Sec.~\ref{transport_prop_sec}.


\section{Linear and nonlinear transport properties}
\label{transport_prop_sec}

The multivariate fluctuation relation~\eqref{sym-Q-magn} generates a hierarchy of time-reversal symmetry relations between the response coefficients and the cumulants of the random currents in the presence of the external magnetic field $\boldsymbol{B}$. Remarkably, such relations can be systematically analyzed at an arbitrary order in the nonequilibrium constraints, as shown in Refs.~\cite{BG19,BG20}. Here, we consider these relations at low orders to apply them thereafter to Aharonov-Bohm rings in Sec.~\ref{AB_sec}. The cumulants of lowest orders and their responses to the affinities are defined in Subsec.~\ref{cumulant_coeff_subsec} below. We then discuss in Subsec.~\ref{galvano_prop_subsec} the particular relations satisfied by these quantities as a consequence of microreversibility. Finally, we show in Subsec.~\ref{comp_sec} how the expression~\eqref{Q_expr} of the cumulant generating function $Q$ can be adequately used in order to compute these cumulants.

\subsection{Cumulants and response coefficients}\label{cumulant_coeff_subsec}

The first, second, and higher cumulants are defined as
\bea
J_{\alpha}(\boldsymbol{A};\boldsymbol{B}) &=& \left. \frac{\partial Q(\boldsymbol{\lambda},\boldsymbol{A};\boldsymbol{B})}{\partial \lambda_{\alpha}} \right\rvert_{\boldsymbol{\lambda}={\boldsymbol{0}}}\, , \label{av_J} \\[0.25cm]
D_{\alpha \beta}(\boldsymbol{A};\boldsymbol{B}) &=& - \frac{1}{2} \left. \frac{\partial^2 Q(\boldsymbol{\lambda},\boldsymbol{A};\boldsymbol{B})}{\partial \lambda_{\alpha} \partial \lambda_{\beta}} \right\rvert_{\boldsymbol{\lambda}={\boldsymbol{0}}} \, , \label{D}\\[0.25cm]
C_{\alpha \beta \gamma}(\boldsymbol{A};\boldsymbol{B}) &=& \left. \frac{\partial^3 Q(\boldsymbol{\lambda},\boldsymbol{A};\boldsymbol{B})}{\partial
\lambda_{\alpha} \partial \lambda_{\beta} \partial \lambda_{\gamma}} \right\rvert_{\boldsymbol{\lambda}={\boldsymbol{0}}} \, , \label{C}\\[0.25cm]
B_{\alpha \beta \gamma \delta}(\boldsymbol{A};\boldsymbol{B}) &=& - \frac{1}{2} \left. \frac{\partial^4 Q(\boldsymbol{\lambda},\boldsymbol{A};\boldsymbol{B})}{\partial \lambda_{\alpha} \partial \lambda_{\beta} \partial \lambda_{\gamma} \partial \lambda_{\delta}} \right\rvert_{\boldsymbol{\lambda}={\boldsymbol{0}}} \, , \label{B}\\[0.25cm]
&\vdots& \nonumber
\eea
Here and in the sequel, the Greek characters $\alpha, \beta, \gamma$ and $\delta$ take any value between $1$ and $2(r-1)$. We hence emphasize that the counting parameter $\lambda_{\alpha}$ can, depending on the actual value of the subscript $\alpha$, either be a thermal counting parameter $\lambda_{jE}$ (for $\alpha = 1, \ldots , r-1$) or a chemical counting parameter $\lambda_{jN}$ [for $\alpha = r, \ldots , 2(r-1)$].

The first cumulants $J_{\alpha}$ merely correspond to the mean currents. They are in general nonlinear functions of the affinities, which are vanishing together at equilibrium.  The response coefficients are introduced by expanding the mean currents in powers of the affinities as
\begin{multline}
J_{\alpha}(\boldsymbol{A};\boldsymbol{B}) = \sum_{\beta} L_{\alpha,\beta}(\boldsymbol{B}) \, A_{\beta} + \frac{1}{2} \, \sum_{\beta,\gamma} M_{\alpha,\beta\gamma}(\boldsymbol{B}) \, A_{\beta} \, A_{\gamma} \\
+ \frac{1}{6} \, \sum_{\beta,\gamma,\delta} N_{\alpha,\beta\gamma\delta}(\boldsymbol{B}) \, A_{\beta} \, A_{\gamma} \, A_{\delta} + \cdots \, ,
\label{Taylor}
\end{multline}
where the coefficients $L_{\alpha,\beta}$ characterize the linear response of the system to the nonequilibrium constraints, while the quantities $M_{\alpha,\beta \gamma}, N_{\alpha,\beta \gamma \delta}, \ldots$ describe the nonlinear response. 

Similarly, the second cumulants $D_{\alpha \beta}$, commonly referred to as the diffusivities, are in general nonlinear functions of $\boldsymbol{A}$ as well. The coefficients $\partial D_{\alpha\beta} / \partial A_{\gamma}, \ldots$ of the power series expansion of $D_{\alpha \beta}$ then define the responses of the diffusivities to the affinities. The same applies to the higher order cumulants $C_{\alpha \beta \gamma}, B_{\alpha \beta \gamma \delta}, \ldots$

\vskip 0.5 cm

\subsection{Time-reversal symmetry relations for cumulants and response coefficients}
\label{galvano_prop_subsec}

Differentiating twice the symmetry relation~\eqref{sym-Q-magn} with respect to $\boldsymbol{\lambda}$ or $\boldsymbol{A}$ yields the Onsager-Casimir reciprocity relations, namely
\be
L_{\alpha,\beta}(\boldsymbol{B}) = L_{\beta,\alpha}(-\boldsymbol{B}) \, ,
\label{Casimir}
\ee
as well as
\be
D_{\alpha\beta}({\boldsymbol{0}};\boldsymbol{B}) = \frac{1}{2} \left[ L_{\alpha , \beta}(\boldsymbol{B}) + L_{\alpha , \beta}(-\boldsymbol{B}) \right] =D_{\alpha\beta}({\boldsymbol{0}};-\boldsymbol{B}) \, ,
\label{FDT}
\ee
for the diffusivities.

Similar relations valid beyond the linear response regime can also be obtained. Taking third derivatives of the symmetry relation~\eqref{sym-Q-magn} leads to the following formulas:
\begin{widetext}
\bea
&&C_{\alpha\beta\gamma}({\boldsymbol{0}};\boldsymbol{B}) = - C_{\alpha\beta\gamma}({\boldsymbol{0}};-\boldsymbol{B}) \; , \label{C0-C0}\\
&&C_{\alpha\beta\gamma}({\boldsymbol{0}};\boldsymbol{B}) = 2\, 
\frac{\partial D_{\alpha\beta}}{\partial A_{\gamma}}({\boldsymbol{0}};\boldsymbol{B}) -2 \, 
\frac{\partial D_{\alpha\beta}}{\partial A_{\gamma}}({\boldsymbol{0}};-\boldsymbol{B}) \; ,
\label{C0-D1}\\
&& M_{\alpha,\beta\gamma}(\boldsymbol{B})+M_{\alpha,\beta\gamma}(-\boldsymbol{B}) = 2\, \frac{\partial D_{\alpha\beta}}{\partial A_{\gamma}}({\boldsymbol{0}};\boldsymbol{B}) +2\,
\frac{\partial D_{\alpha\gamma}}{\partial A_{\beta}}({\boldsymbol{0}};-\boldsymbol{B}) \; , \label{M-D1}\\
&& C_{\alpha\beta\gamma}({\boldsymbol{0}};\boldsymbol{B}) = -\left(M_{\alpha,\beta\gamma}+M_{\beta,\gamma\alpha}+M_{\gamma,\alpha\beta}\right)_{\boldsymbol{B}}
+ 2 \left(\frac{\partial D_{\alpha\beta}}{\partial A_{\gamma}} 
+ \frac{\partial D_{\beta\gamma}}{\partial A_{\alpha}}
+ \frac{\partial D_{\gamma\alpha}}{\partial A_{\beta}}\right)_{\boldsymbol{A}={\boldsymbol{0}};\,\boldsymbol{B}} \; .
\label{M-D1-C0}
\eea
\end{widetext}
These formulas correspond respectively to Eqs.~(20), (22), (23), and~(24) of Ref.~\cite{BG19} and they confirm results obtained in Ref.~\cite{Gas13NJP}.

The third cumulants~\eqref{C} characterize the magnetic-field asymmetry of the fluctuations \cite{SB04,FB08}. At equilibrium where $\boldsymbol{A}={\boldsymbol{0}}$, they are odd with respect to the magnetic field because of Eq.~\eqref{C0-C0}, so that this magnetic-field asymmetry disappears in the absence of a magnetic field, i.e., for $\boldsymbol{B}={\boldsymbol{0}}$, and the third cumulants then vanish at equilibrium, $C_{\alpha\beta\gamma}({\boldsymbol{0}};{\boldsymbol{0}})=0$.  Moreover, Eq.~\eqref{C0-D1} shows that the third cumulants can be expressed in terms of the first responses of the diffusivities~\eqref{D} with respect to the affinities. Furthermore, the third cumulants are fully given in terms of the second response coefficients according to
\begin{multline}
C_{\alpha\beta\gamma}({\boldsymbol{0}};\boldsymbol{B}) = \left(M_{\alpha,\beta\gamma}+M_{\beta,\gamma\alpha}+M_{\gamma,\alpha\beta}\right)_{\boldsymbol{B}} \\
-\left(M_{\alpha,\beta\gamma}+M_{\beta,\gamma\alpha}+M_{\gamma,\alpha\beta}\right)_{-\boldsymbol{B}} \; ,
\label{C-M-odd}
\end{multline}
which is deduced using Eq.~\eqref{M-D1-C0} for $\pm\boldsymbol{B}$ combined with Eqs.~\eqref{C0-C0} and~\eqref{C0-D1}.

If $\alpha=\beta=\gamma$, we recover relations obtained in Ref.~\cite{SU08} giving the unidirectional third cumulants and the sensitivity of the diffusivities in terms of the unidirectional response coefficients, i.e.,
\be
C_{\alpha\alpha\alpha}({\boldsymbol{0}};\boldsymbol{B}) = 3\left[ M_{\alpha,\alpha\alpha}(\boldsymbol{B})-M_{\alpha,\alpha\alpha}(-\boldsymbol{B})\right] \; ,
\label{C-M-odd-iii}
\ee
and
\be
\frac{\partial D_{\alpha\alpha}}{\partial A_{\alpha}}({\boldsymbol{0}};\boldsymbol{B})  = 2 \, M_{\alpha,\alpha\alpha}(\boldsymbol{B})-M_{\alpha,\alpha\alpha}(-\boldsymbol{B})\; .
\label{D-M-iii}
\ee
Reciprocally, Eq.~\eqref{M-D1-C0} with $\alpha=\beta=\gamma$ implies that the unidirectional response coefficients are related to the previous quantities as
\be
M_{\alpha,\alpha\alpha}(\boldsymbol{B}) = 2\, \frac{\partial D_{\alpha\alpha}}{\partial A_{\alpha}}({\boldsymbol{0}};\,\boldsymbol{B}) - \frac{1}{3} \, C_{\alpha\alpha\alpha}({\boldsymbol{0}};\,\boldsymbol{B}) \, .
\label{M-D-C-iii}
\ee
Equations~\eqref{M-D1}-\eqref{C-M-odd} concern the general case where $\alpha\neq\beta\neq\gamma$.

From the fourth derivatives of the symmetry relation~\eqref{sym-Q-magn}, we similarly deduce that
\begin{widetext}
\bea
&& B_{\alpha\beta\gamma\delta}({\boldsymbol{0}};\boldsymbol{B}) = B_{\alpha\beta\gamma\delta}({\boldsymbol{0}};-\boldsymbol{B}) \; , \label{B0-B0}\\
&& B_{\alpha\beta\gamma\delta}({\boldsymbol{0}};\boldsymbol{B}) = \frac{1}{2} \, \frac{\partial C_{\alpha\beta\gamma}}{\partial A_{\delta}}({\boldsymbol{0}};\boldsymbol{B}) + \frac{1}{2} \, \frac{\partial C_{\alpha\beta\gamma}}{\partial A_{\delta}}({\boldsymbol{0}};-\boldsymbol{B})\; , \label{B0-C1}\\
&& \frac{\partial^2 D_{\alpha\beta}}{\partial A_{\gamma}\partial A_{\delta}}({\boldsymbol{0}};\boldsymbol{B}) 
-  \frac{1}{2} \, \frac{\partial C_{\alpha\beta\gamma}}{\partial A_{\delta}}({\boldsymbol{0}};\boldsymbol{B}) 
= \frac{\partial^2 D_{\alpha\beta}}{\partial A_{\gamma}\partial A_{\delta}}({\boldsymbol{0}};-\boldsymbol{B})  
- \frac{1}{2} \, \frac{\partial C_{\alpha\beta\delta}}{\partial A_{\gamma}}({\boldsymbol{0}};-\boldsymbol{B})\; , \label{D2-C1}\\
&& B_{\alpha\beta\gamma\delta}({\boldsymbol{0}};\boldsymbol{B}) = \frac{1}{2}\, N_{\alpha,\beta\gamma\delta}(\boldsymbol{B})+\frac{1}{2}\, N_{\alpha,\beta\gamma\delta}(-\boldsymbol{B}) \nonumber\\
&&\qquad\qquad\qquad\qquad
-\left( \frac{\partial^2 D_{\alpha\beta}}{\partial A_{\gamma}\partial A_{\delta}} 
+ \frac{\partial^2 D_{\alpha\gamma}}{\partial A_{\beta}\partial A_{\delta}} 
+ \frac{\partial^2 D_{\alpha\delta}}{\partial A_{\beta}\partial A_{\gamma}} \right)_{\boldsymbol{A}={\boldsymbol{0}};\,\boldsymbol{B}} \nonumber\\
&&\qquad\qquad\qquad\quad\qquad\qquad
+\frac{1}{2}\left(\frac{\partial C_{\alpha\beta\gamma}}{\partial A_{\delta}}
+\frac{\partial C_{\alpha\beta\delta}}{\partial A_{\gamma}}
+\frac{\partial C_{\alpha\gamma\delta}}{\partial A_{\beta}}\right)_{\boldsymbol{A}={\boldsymbol{0}};\,\boldsymbol{B}} \; , \label{N-N}\\
&& B_{\alpha\beta\gamma\delta}({\boldsymbol{0}};\boldsymbol{B}) = \frac{1}{2}\, \left(N_{\alpha,\beta\gamma\delta}+N_{\beta,\gamma\delta\alpha}+N_{\gamma,\delta\alpha\beta} +N_{\delta,\alpha\beta\gamma}\right)_{\boldsymbol{B}} \nonumber\\
&&\qquad\qquad\qquad\qquad
 - \left( \frac{\partial^2 D_{\alpha\beta}}{\partial A_{\gamma}\partial A_{\delta}} 
+ \frac{\partial^2 D_{\alpha\gamma}}{\partial A_{\beta}\partial A_{\delta}} 
+ \frac{\partial^2 D_{\alpha\delta}}{\partial A_{\beta}\partial A_{\gamma}} 
+ \frac{\partial^2 D_{\beta\gamma}}{\partial A_{\alpha}\partial A_{\delta}} 
+ \frac{\partial^2 D_{\beta\delta}}{\partial A_{\alpha}\partial A_{\gamma}} 
+ \frac{\partial^2 D_{\gamma\delta}}{\partial A_{\alpha}\partial A_{\beta}}\right)_{\boldsymbol{A}={\boldsymbol{0}};\,\boldsymbol{B}}
\nonumber\\
&& \qquad\qquad\qquad\qquad\qquad
+ \frac{1}{2} \left(
\frac{\partial C_{\beta\gamma\delta}}{\partial A_{\alpha}}
+\frac{\partial C_{\gamma\delta\alpha}}{\partial A_{\beta}} 
+\frac{\partial C_{\delta\alpha\beta}}{\partial A_{\gamma}}
+\frac{\partial C_{\alpha\beta\gamma}}{\partial A_{\delta}}\right)_{\boldsymbol{A}={\boldsymbol{0}};\,\boldsymbol{B}} \; . 
\label{N-D2-C1-B0}
\eea
These formulas correspond respectively to Eqs.~(31), (32), (33), (35), and~(36) of Ref.~\cite{BG19}.

In the absence of a magnetic field, $\boldsymbol{B}={\boldsymbol{0}}$, these relations simplify to \cite{AG04,AG07JSM}
\bea
&& B_{\alpha\beta\gamma\delta}({\bf 0};{\bf 0}) = \left(\frac{\partial C_{\alpha\beta\gamma}}{\partial A_{\delta}} \right)_{\boldsymbol{A}={\boldsymbol{0}};\,\boldsymbol{B}={\boldsymbol{0}}} \; ,\\
&& N_{\alpha,\beta\gamma\delta}({\bf 0})
= \left( \frac{\partial^2 D_{\alpha\beta}}{\partial A_{\gamma}\partial A_{\delta}} 
+\frac{\partial^2 D_{\alpha\gamma}}{\partial A_{\beta}\partial A_{\delta}} 
+ \frac{\partial^2 D_{\alpha\delta}}{\partial A_{\beta}\partial A_{\gamma}} - \frac{1}{2}\, B_{\alpha\beta\gamma\delta}\right)_{\boldsymbol{A}={\boldsymbol{0}};\,\boldsymbol{B}={\boldsymbol{0}}} \; .
\eea
\end{widetext}

Similar relations can also be deduced at higher orders. A detailed analysis of such relations at an arbitrary order has been recently performed in Ref.~\cite{BG18} in the absence of a magnetic field. This has then been generalized in Refs.~\cite{BG19,BG20} to the case of a nonzero magnetic field $\boldsymbol{B}$.

The multivariate fluctuation relation~\eqref{sym-Q-magn} hence generates relations between the cumulants and their responses to the nonequilibrium constraints. The dependence of these quantities on the magnetic field $\boldsymbol{B}$ is thus considerably constrained by microreversibility. These results fully rely on the time-reversal symmetry expressed in Eq.~\eqref{sym-Q-magn}. In particular, no reference has been made to the actual functional form of the cumulant generating function $Q(\boldsymbol{\lambda},\boldsymbol{A};\boldsymbol{B})$. We now discuss how the expression~\eqref{Q_expr} of $Q$ in terms of the scattering unitary matrix $\hat S(\varepsilon;\boldsymbol{B})$ can be used in order to compute the cumulants and their responses.


\subsection{Computing the cumulants from their generating function}
\label{comp_sec}

In order to obtain the cumulants, we write the cumulant generating function~\eqref{Q_expr} as the trace \cite{Gas15NJP}
\be
Q(\boldsymbol{\lambda},\boldsymbol{A};\boldsymbol{B}) = - g_s \int \frac{d\varepsilon}{2\pi\hbar} \, {\rm tr}\ln\left[1+ \hat R(\varepsilon;\boldsymbol{\lambda},\boldsymbol{A};\boldsymbol{B})\right]
\label{Q-LL-2}
\ee
with
\begin{multline}
\hat R(\varepsilon;\boldsymbol{\lambda},\boldsymbol{A};\boldsymbol{B})\equiv \hat f(\varepsilon) \\
\times \left[\hat S^{\dagger}(\varepsilon;\boldsymbol{B}) \, {\rm e}^{\varepsilon\, \hat\lambda_E + \hat \lambda_N} \, \hat S(\varepsilon;\boldsymbol{B}) \, {\rm e}^{-\varepsilon\, \hat\lambda_E - \hat \lambda_N} - 1 \right] \, .
\label{R_def}
\end{multline}
The dependence of the quantity~\eqref{R_def} on the affinities $\boldsymbol{A}$ given by Eqs.~\eqref{A_E}-\eqref{A_def} finds its origin into the diagonal matrix $\hat f(\varepsilon)$, whose elements are the Fermi-Dirac distributions~\eqref{Fermi_Dirac_dis} of the reservoirs. On the other hand, the dependence of the quantity $\hat R$ on the counting parameters $\boldsymbol{\lambda}$ defined by Eq.~\eqref{lambda_def} rises from the matrices $\exp(\pm \varepsilon\, \hat\lambda_E \pm \hat \lambda_N)$.

We now take successive derivatives of the cumulant generating function $Q(\boldsymbol{\lambda},\boldsymbol{A};\boldsymbol{B})$ with respect to the counting parameters $\lambda_{\alpha}, \lambda_{\beta} , \ldots$ We recall (see beginning of Subsec.~\ref{cumulant_coeff_subsec} above) that the Greek characters $\alpha, \beta, \ldots$ take values between $1$ and $2(r-1)$. We hence have, upon differentiating Eq.~\eqref{Q-LL-2},
\begin{widetext}
\bea
\partial_{\alpha} Q &=& - g_s \int \frac{d\varepsilon}{2 \pi \hbar} \, {\rm tr} \left( \partial_{\alpha} \hat R \; \frac{1}{1+\hat R} \right)  , \label{d_alpha_Q} \\
\partial_{\alpha} \partial_{\beta} Q &=& - g_s \int \frac{d\varepsilon}{2 \pi \hbar} \, {\rm tr} \left( \partial_{\alpha} \partial_{\beta} \hat R \; \frac{1}{1+\hat R} + \partial_{\alpha} \hat R \; \partial_{\beta} \frac{1}{1+\hat R} \right)  , \label{d_alpha_beta_Q} \\
\partial_{\alpha} \partial_{\beta} \partial_{\gamma} Q &=& - g_s \int \frac{d\varepsilon}{2 \pi \hbar} \, {\rm tr} \left( \partial_{\alpha} \partial_{\beta} \partial_{\gamma} \hat R \; \frac{1}{1+\hat R} + \partial_{\alpha} \partial_{\beta} \hat R \; \partial_{\gamma} \frac{1}{1+\hat R} + \partial_{\alpha} \partial_{\gamma} \hat R \; \partial_{\beta} \frac{1}{1+\hat R} + \partial_{\alpha} \hat R \; \partial_{\beta} \partial_{\gamma} \frac{1}{1+\hat R} \right)  , \label{d_alpha_beta_gamma_Q} \\
& \vdots & \nonumber
\eea
\end{widetext}
with the notations $\partial_{\alpha} \equiv \partial / \partial\lambda_{\alpha}, \ldots$ and the convention that these partial derivatives only apply to the operator $\hat R$ or $(1+\hat R)^{-1}$ (and possibly derivatives of them) immediately on its right-hand side. It is worth noting that the quantities $\partial_{\alpha} Q, \, \partial_{\alpha} \partial_{\beta} Q, \, \partial_{\alpha} \partial_{\beta} \partial_{\gamma} Q, \ldots$ are invariant under any permutation of the indices $\alpha, \beta, \gamma, \ldots$ of the derivatives that act on the cumulant generating function $Q$. This can be explicitly checked from the power series expression $(1+\hat R)^{-1} = \sum_{n=0}^{\infty} (-1)^n {\hat R}^n$. Now, we have that
\begin{widetext}
\bea
\left. \partial_{\alpha} \frac{1}{1+\hat R} \right\rvert_{\boldsymbol{\lambda}={\boldsymbol{0}}} &=& - \hat R_{\alpha} \, , \label{d_alpha_1_R}\\
\left. \partial_{\alpha} \partial_{\beta} \frac{1}{1+\hat R} \right\rvert_{\boldsymbol{\lambda}={\boldsymbol{0}}} &=& - \hat R_{\alpha \beta} + \hat R_{\alpha}\hat R_{\beta} + \hat R_{\beta} \hat R_{\alpha} \, , \label{d_alpha_beta_1_R}\\
\left. \partial_{\alpha} \partial_{\beta} \partial_{\gamma} \frac{1}{1+\hat R} \right\rvert_{\boldsymbol{\lambda}={\boldsymbol{0}}} &=& - \hat R_{\alpha \beta \gamma} + \hat R_{\alpha \beta} \hat R_{\gamma} + \hat R_{\alpha \gamma} \hat R_{\beta}
+ \hat R_{\alpha} \hat R_{\beta \gamma} + \hat R_{\beta \gamma} \hat R_{\alpha} + \hat R_{\beta} \hat R_{\alpha \gamma} + \hat R_{\gamma} \hat R_{\alpha \beta} \nonumber \\
&& - \hat R_{\alpha} \hat R_{\beta} \hat R_{\gamma} - \hat R_{\alpha} \hat R_{\gamma} \hat R_{\beta} - \hat R_{\beta} \hat R_{\alpha} \hat R_{\gamma} - \hat R_{\gamma} \hat R_{\alpha} \hat R_{\beta} - \hat R_{\beta} \hat R_{\gamma} \hat R_{\alpha} - \hat R_{\gamma} \hat R_{\beta} \hat R_{\alpha} \, , \label{d_alpha_beta_gamma_1_R}\\
& \vdots & \nonumber
\eea
with the notations
\bea
\hat R_{\alpha} \equiv \left. \partial_{\alpha} \hat R \right\rvert_{\boldsymbol{\lambda}={\boldsymbol{0}}} &=& \varepsilon^{\nu_{\alpha}} \hat f \left( \hat S^{\dagger} \hat P_{\kappa_{\alpha}} \hat S - \hat P_{\kappa_{\alpha}} \right) \, , \label{R1}\\
\hat R_{\alpha \beta} \equiv \left. \partial_{\alpha} \partial_{\beta} \hat R \right\rvert_{\boldsymbol{\lambda}={\boldsymbol{0}}} &=& \varepsilon^{\nu_{\alpha}} \varepsilon^{\nu_{\beta}} \hat f \left[ \delta_{\alpha \beta} \left( \hat S^{\dagger} \hat P_{\kappa_{\alpha}} \hat S + \hat P_{\kappa_{\alpha}} \right) - \hat S^{\dagger} \hat P_{\kappa_{\alpha}} \hat S \hat P_{\kappa_{\beta}} - \hat S^{\dagger} \hat P_{\kappa_{\beta}} \hat S \hat P_{\kappa_{\alpha}} \right] \, , \label{R2}\\
\hat R_{\alpha \beta \gamma} \equiv \left. \partial_{\alpha} \partial_{\beta} \partial_{\gamma} \hat R \right\rvert_{\boldsymbol{\lambda}={\boldsymbol{0}}} &=& \varepsilon^{\nu_{\alpha}} \varepsilon^{\nu_{\beta}} \varepsilon^{\nu_{\gamma}} \hat f  \left[ \delta_{\alpha \beta} \delta_{\alpha \gamma} \left( \hat S^{\dagger} \hat P_{\kappa_{\alpha}} \hat S - \hat P_{\kappa_{\alpha}} \right) - \delta_{\alpha \beta} \left( \hat S^{\dagger} \hat P_{\kappa_{\alpha}} \hat S \hat P_{\kappa_{\gamma}} - \hat S^{\dagger} \hat P_{\kappa_{\gamma}} \hat S \hat P_{\kappa_{\alpha}} \right) \right. \nonumber\\
&& \left. - \delta_{\alpha \gamma} \left( \hat S^{\dagger} \hat P_{\kappa_{\alpha}} \hat S \hat P_{\kappa_{\beta}} - \hat S^{\dagger} \hat P_{\kappa_{\beta}} \hat S \hat P_{\kappa_{\alpha}} \right) - \delta_{\beta \gamma} \left( \hat S^{\dagger} \hat P_{\kappa_{\beta}} \hat S \hat P_{\kappa_{\alpha}} - \hat S^{\dagger} \hat P_{\kappa_{\alpha}} \hat S \hat P_{\kappa_{\beta}} \right) \right] \, , \label{R3}\\
&\vdots&\nonumber
\eea
where the quantities $\nu_{\alpha}$ and $\kappa_{\alpha}$ are defined by
\end{widetext}
\be
\nu_{\alpha} \equiv \left\{\begin{array}{ll}
1 \, , &\qquad \text{if $\alpha = 1, \ldots , r-1$,} \\[0.15cm]
0 \, , & \qquad \text{if $\alpha = r, \ldots , 2(r-1)$,}
\end{array} \right.
\label{nu_alpha_def}
\ee
and
\be
\kappa_{\alpha} \equiv \left\{\begin{array}{ll}
\alpha \, , &\qquad \text{if $\alpha = 1, \ldots , r-1$,} \\[0.15cm]
\alpha - r + 1 \, , &\qquad \text{if $\alpha = r, \ldots , 2(r-1)$,}
\end{array} \right. 
\label{kappa_alpha_def}
\ee
while $\hat P_{\kappa_{\alpha}}$ is the projector on the reservoir $\kappa_{\alpha}$ defined by the $r \times r$ matrix with a single non-vanishing element equal to one on the diagonal at the $\kappa_{\alpha}^{\rm th}$ row and $\kappa_{\alpha}^{\rm th}$ column. Indeed, note that, in view of its definition~\eqref{kappa_alpha_def}, the integer $\kappa_{\alpha}$ takes values between 1 and $r-1$ for any value of $\alpha$ [$\alpha = 1, \ldots , 2(r-1)$]. Therefore, combining Eqs.~\eqref{d_alpha_Q}-\eqref{d_alpha_beta_gamma_Q} with Eqs.~\eqref{d_alpha_1_R}-\eqref{d_alpha_beta_gamma_1_R}, we find that
\begin{widetext}
\bea
\left. \frac{\partial Q}{\partial \lambda_{\alpha}} \right\rvert_{\boldsymbol{\lambda}={\boldsymbol{0}}} &=& - g_s \int \frac{d\varepsilon}{2 \pi \hbar} \, {\rm tr}\, \hat R_{\alpha} \, , \label{Q1} \\
\left. \frac{\partial^2 Q}{\partial \lambda_{\alpha} \partial \lambda_{\beta}}\right\rvert_{\boldsymbol{\lambda}={\boldsymbol{0}}} &=& - g_s \int \frac{d\varepsilon}{2 \pi \hbar} \, {\rm tr} \left( \hat R_{\alpha \beta} - \hat R_{\alpha} \hat R_{\beta} \right) \, , \label{Q2}\\
\left. \frac{\partial^3 Q}{\partial \lambda_{\alpha} \partial \lambda_{\beta} \partial \lambda_{\gamma}} \right\rvert_{\boldsymbol{\lambda}={\boldsymbol{0}}} &=& - g_s \int \frac{d\varepsilon}{2 \pi \hbar} \, {\rm tr} \left( \hat R_{\alpha \beta \gamma} - \hat R_{\alpha \beta} \hat R_{\gamma} - \hat R_{\alpha \gamma} \hat R_{\beta} - \hat R_{\alpha} \hat R_{\beta \gamma} + \hat R_{\alpha} \hat R_{\beta} \hat R_{\gamma} + \hat R_{\alpha} \hat R_{\gamma} \hat R_{\beta} \right) \, , \label{Q3}\\
&\vdots& \nonumber
\eea
\end{widetext}

At equilibrium, all the reservoirs share the same temperature and chemical potential, so that the affinities~\eqref{A_E}-\eqref{A_N} are all equal to zero, i.e., $\boldsymbol{A}={\boldsymbol{0}}$. In this case, the matrix of Fermi-Dirac distributions is proportional to the identity matrix, e.g., $\hat f = f_r \, \hat 1$ if the reservoirs all have the temperature $T_r$ and chemical potential $\mu_r$ of the $r^{\rm th}$ reservoir. Using the unitarity of the $S$-matrix, Eqs.~\eqref{R1} and~\eqref{Q1} imply that the mean currents~\eqref{av_J} are vanishing at equilibrium, i.e., $J_{\alpha} \vert_{\boldsymbol{A}={\boldsymbol{0}}} = 0$, as expected.  Similarly, Eqs.~\eqref{R1}-\eqref{R3} and~\eqref{Q3} imply that the third cumulant~\eqref{C} for a single current is also vanishing at equilibrium, i.e., $C_{\alpha\alpha\alpha} \vert_{\boldsymbol{A}={\boldsymbol{0}}} = 0$.

If the system is at equilibrium and furthermore at zero absolute temperature, the equilibrium Fermi-Dirac distribution is given by the Heaviside step function $f_r(\varepsilon)=\theta(\mu_{r}-\varepsilon)$, where $\mu_{r}$ is the common chemical potential of all the reservoirs, so that $\hat f(\varepsilon)=\hat 1$ if $\varepsilon <\mu_r$ and $\hat f(\varepsilon)=0$ if $\varepsilon >\mu_r$, giving ${\rm tr}\ln(1+ \hat R)=0$ under both conditions in Eq.~\eqref{Q-LL-2} because of Eq.~\eqref{R_def}. In this case, the cumulant generating function~\eqref{Q-LL-2} is identically equal to zero, $Q(\boldsymbol{\lambda},{\boldsymbol{0}};\boldsymbol{B})_{T=0} =0$, so that all the cumulants are vanishing and there is no fluctuation at zero absolute temperature.


\section{Application to Aharonov-Bohm rings}
\label{AB_sec}

The above considerations are valid for a generic multi-terminal circuit. In the present section, we illustrate them on a particular system where the scattering region consists in an Aharonov-Bohm ring. We describe the general setup in Subsec.~\ref{AB_set_up_subsec}, before we make a detailed analysis of the specific cases where the ring is connected to 2, 3 and 4 terminals in Subsecs.~\ref{AB_2_term_subsec},~\ref{AB_3_term_subsec}, and~\ref{AB_4_term_subsec}, respectively. We emphasize that, throughout this section, we consider an isothermal system, which means that all reservoirs share the same temperature $T$ but still have different chemical potentials. Therefore, only currents of electrons flow within the circuit. The counting parameters $\boldsymbol{\lambda}$ and affinities $\boldsymbol{A}$ hence merely have $r-1$ components, and they are here given by
\be
\boldsymbol{\lambda} = \left( \lambda_{1N} , \ldots , \lambda_{r-1 , N} \right)
\label{isothermal_lambda_def}
\ee
and
\be
\boldsymbol{A} = \left( A_{1N} , \ldots , A_{r-1 , N} \right) \, ,
\label{isothermal_A_def}
\ee
where $A_{jN} = \left( \mu_j - \mu_r \right)/(k_{\rm B} T)$ in view of Eq.~\eqref{A_N}.


\subsection{General set up}\label{AB_set_up_subsec}

We consider a circular Aharonov-Bohm ring of radius $R$ connected to $r$ terminals, with a static magnetic field $\boldsymbol{B} = B \boldsymbol{u}_z$ perpendicular to the plane of the ring. We have $B \neq 0$ only inside the ring. The magnetic flux through the section area $\Sigma = \pi R^2$ enclosed by the ring is equal to $\Phi = B \Sigma$. The corresponding dimensionless magnetic flux is denoted by
\be
\phi=\frac{e}{\hbar}\, \Phi = \frac{e}{\hbar}\, B \, \Sigma \, ,
\label{rescaled_mag_flux}
\ee
which will be taken to assume values between $-\pi$ and $\pi$ in Subsecs.~\ref{AB_2_term_subsec}-\ref{AB_4_term_subsec}. The vector potential $\boldsymbol{\mathcal{A}} \left( {\bf r} ; \boldsymbol{B} \right)$ associated with the magnetic field $\boldsymbol{B}$ is oriented along the orthoradial unit vector $\boldsymbol{u}_{\theta}$ of the cylindrical coordinates (adapted to the description of the ring), and we have
\be
\boldsymbol{\mathcal{A}} \left( {\bf r} ; \boldsymbol{B} \right) = \mathcal{A}_{\theta} \, \boldsymbol{u}_{\theta} = \frac{\Phi}{2 \pi R} \, \boldsymbol{u}_{\theta} \, .
\label{pot_vector_AB}
\ee
The wires forming the ring and the terminals are supposed to be one-dimensional. Following the discussion of Subsec.~\ref{multi_term_subsec}, this assumption for the terminals can be understood as representing the limit case of a waveguide that has a very narrow spatial extension in the transverse directions. Substituting the expression~\eqref{pot_vector_AB} of the vector potential into the expression~\eqref{one_body_Ham_split} of the Hamiltonian operator $\hat h_{\mathrm{scat}}(\boldsymbol{B})$ [where we take the potential $u({\bf r})$ to be zero] and writing the Laplacian in cylindrical coordinates, we see that the Hamiltonian operator $\hat h_{\rm{ring}}$ on the ring takes the form
\be
\hat h_{\rm{ring}} = -\frac{\hbar^2}{2mR^2}\left(\frac{d}{d\theta} + i \, \frac{\Phi}{\Phi_0}\right)^2 \, ,
\label{Ham_theta}
\ee
with $\theta$ the polar angle of the cylindrical coordinates and the quantum magnetic flux
\be
\Phi_0 = 2\pi \, \frac{\hbar}{e} \, .
\ee
Note that both quantities $\Phi$ and $\Phi_0$ are independent of the variable $\theta$. We now change the independent variable describing the ring from the polar angle $\theta$ to the position $x$ along the ring, with $x = R \theta$. The Hamiltonian operator~\eqref{Ham_theta} hence reads
\be
\hat h_{\rm{ring}} = -\frac{\hbar^2}{2m}\left(\frac{d}{dx} + i \, f\right)^2 \, ,
\label{Ham_ring}
\ee
where the quantity $f$ is related to the fluxes $\Phi$ and $\Phi_0$ through
\be
f \equiv \frac{\Phi}{R\Phi_0} = \frac{\phi}{2\pi R} \, .
\label{f_def}
\ee
In the terminal $j$, the Hamiltonian operator $\hat h_{j}$ is merely the free-particle one, that is
\be
\hat h_{j} = -\frac{\hbar^2}{2m}\frac{d^2}{dx^2} \, ,
\label{Ham_term_AB}
\ee
where the $x$-axis is oriented toward infinity. As was already discussed in Subsec.~\ref{multi_term_subsec}, $x$ denotes the position along each of the $r$ terminals. The latter are thus semi-infinite wires extending from $x=0$ to $x=\infty$ and connected to the ring at some angles $\theta_j$, $j=1,...,r$.  In conclusion, the Aharonov-Bohm Hamiltonian operator $\hat h_{\rm{AB}}$ that describes the circuit considered here is precisely of the form~\eqref{one_body_Ham_split}, and we have
\be
\hat h_{\rm{AB}} = \left\{\begin{array}{ll}
\hat h_{\rm{ring}} = -\frac{\hbar^2}{2m}\left(\frac{d}{dx} + i \, f\right)^2 \, , &\ \text{on the ring,} \\[0.3cm]
\hat h_{j} = -\frac{\hbar^2}{2m}\frac{d^2}{dx^2} \, , &\ \text{in the terminal $j$.}
\end{array} \right. 
\label{Ham_AB_split}
\ee
We now solve the time-independent Schr\"odinger equation
\be
\hat h_{\rm{AB}} \, \psi = \varepsilon \, \psi
\label{Schr_eq_gen}
\ee
separately in each terminal $j=1, \ldots , r$, and on the ring.

The solution of the Schr\"odinger equation $\hat h_j\, \psi_j (x) = \varepsilon\, \psi_j (x)$ is given by
\be
\psi_j(x) = a_j \, {\rm e}^{-ikx} + b_j \, {\rm e}^{ikx} \, ,
\label{term_eigenstates}
\ee
in the terminal $j$ and thus for $x>0$, where the wave number $k$ is related to the energy $\varepsilon$ through $\varepsilon = \hbar^2 k^2 / 2m$. We emphasize that, because we solve the full Schr\"odinger equation~\eqref{Schr_eq_gen}, we must have a fixed energy $\varepsilon$ and thus, by extension, a fixed wave number $k$ over the whole circuit. That is, the wave functions $\psi_j$ have the same wave number $k$ in each reservoir $j$. Now, the solution of the Schr\"odinger equation $\hat h_{\rm{ring}} \psi_{j,j+1}(x) = \varepsilon \psi_{j,j+1}(x)$ is
\be
\psi_{j,j+1}(x) = a_{j,j+1} \, {\rm e}^{i(k-f)x} + b_{j,j+1} \, {\rm e}^{-i(k+f)x} \, ,
\label{ring_eigenstates}
\ee
on the ring between the terminals $j$ and $j+1$, and thus for $R\theta_j < x < R\theta_{j+1}$.  Here again, we emphasize that the wave number $k$ in Eq.~\eqref{ring_eigenstates} is the same as in the terminal eigenstates~\eqref{term_eigenstates}. The expressions~\eqref{ring_eigenstates} are valid for values $j=1, \ldots , r$ under the condition that we identify the value $j=r+1$ to actually mean $j=1$. The wave function of an electron in the entire circuit is thus completely characterized by the $4r$~unknown quantities $\left\{ a_{j} , b_{j} , a_{j,j+1} , b_{j,j+1} \right\}_{j = 1, \ldots , r}$.

At the vertex between the terminal $j$ and the point $x=R\theta_j$ of the ring, the wave function is required to be continuous and to obey Neumann boundary conditions, so that we have
\be
\psi_j(0) = \psi_{j-1,j}(R\theta_j) = \psi_{j,j+1}(R\theta_j)
\label{bc1}
\ee
for the wave function, and
\begin{multline}
\frac{d\psi_j}{dx}(0) - \left( \frac{d}{dx} + if\right)\psi_{j-1,j}(R\theta_j) \\
+ \left( \frac{d}{dx} + if\right)\psi_{j,j+1}(R\theta_j) = 0
\label{bc2}
\end{multline}
for the derivative of the wave function. These conditions hold for $j=1,...,r$ by identifying $j=0$ with $j=r$, and $j=r+1$ with $j=1$. Hence, we have $2r$ relations~\eqref{bc1} and $r$ relations~\eqref{bc2}, that is a total of $3r$ conditions imposed on the wave function. Apparently, $r$ conditions are missing in order to uniquely determine the $4r$ unknowns of the problem. However, such additional conditions are not needed since our aim is to derive the scattering matrix of the Aharonov-Bohm ring connected to the terminals. The rationale of restricting our attention to the scattering matrix rises from the fact that it fully specifies the cumulant generating function $Q$, as is clear on the expression~\eqref{Q_expr} [or equivalently~\eqref{Q-LL-2}] of the latter. We recall that $Q$ expresses the full counting statistics of the electron currents that take place within the circuit. Therefore, the knowledge of the scattering matrix provides us with a complete description of the transport properties of the Aharonov-Bohm circuit considered here. Now, as we already stated at the end of Subsec.~\ref{multi_term_subsec}, the scattering matrix is the $r \times r$ unitary matrix $\left( S_{jj'} \right)_{1 \leqslant j , j' \leqslant r}$ that expresses each coefficient~$b_j$ of the terminal eigenstates~\eqref{term_eigenstates} in terms of their coefficients~$a_j$, that is
\be
b_j = \sum_{j'=1}^{r} S_{jj'} a_{j'} \, ,
\ee
for $j = 1, \ldots , r$. The $3r$ conditions~\eqref{bc1}-\eqref{bc2} are thus precisely sufficient in order to unambiguously obtain the scattering matrix. This is done in Subsecs.~\ref{AB_2_term_subsec},~\ref{AB_3_term_subsec}, and~\ref{AB_4_term_subsec} below in the cases of 2, 3, and 4 terminals, respectively.

The $3r$ linear Eqs.~\eqref{bc1}-\eqref{bc2} can be exactly solved in order to obtain an analytic expression for the scattering matrix. The latter will be seen to be a function $\hat S_{k}(f)$ of $k$ and $f$, i.e., of the energy $\varepsilon$ and the magnetic flux $\Phi$ in view of $\varepsilon = \hbar^2 k^2 / 2 m$ and the expression~\eqref{f_def} defining $f$, respectively. It will of course also depend on the radius $R$ of the ring and on the angles $\theta_j$, but these quantities will be treated as mere parameters. We then use the resulting expression of $\hat S_{k}(f)$ to obtain the cumulants, as well as their responses to the nonequilibrium constraints, by means of numerical evaluations of the integrals~\eqref{Q1}-\eqref{Q3} (in which we make use of the relation $\varepsilon = \hbar^2 k^2 / 2 m$) and of their derivatives with respect to the affinities. This is done for different values of the magnetic flux $\phi$, i.e., of the magnetic field $B$ in view of Eq.~\eqref{rescaled_mag_flux}, which hence allows us to illustrate some of the relations obtained in Sec.~\ref{transport_prop_sec} as a direct consequence of the multivariate fluctuation relation~\eqref{sym-Q-magn}. We emphasize that the time-reversal symmetry relations discussed in Sec.~\ref{transport_prop_sec} involve the cumulants and their responses at equilibrium, i.e., for $\boldsymbol{A}=\boldsymbol{0}$, where the reservoirs all share the same chemical potential $\mu$ (and of course the same temperature $T$ since we consider an isothermal system). We hence compute the integrals~\eqref{Q1}-\eqref{Q3} for values of $T$ and $\mu$ that are the same in each of the reservoirs.  Additional details about the numerical methods used in the computations are given in Appendix~\ref{numerics}.

\vskip 0.5 cm

\subsection{Aharonov-Bohm ring with 2 terminals}\label{AB_2_term_subsec}

\begin{figure}[h]
\begin{center}
\includegraphics[scale=0.35]{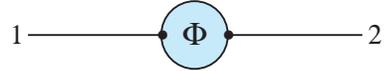}
\end{center}
\caption{Schematic representation of an Aharonov-Bohm ring with 2 terminals forming an angle $\vartheta=\pi$.} 
\label{fig2}
\end{figure}

Let us first consider an Aharonov-Bohm ring connected to two terminals located at the angles $\theta_1=0$ and $\theta_2=\vartheta$, as is illustrated in Fig.~\ref{fig2} in the particular case of an angle $\vartheta = \pi$.

In this case, the scattering matrix is a $2 \times 2$ matrix. As we discussed in Subsec.~\ref{AB_set_up_subsec} above, we use the boundary conditions~\eqref{bc1}-\eqref{bc2} for $j=1,2$ to express the coefficients $\{b_1, b_2\}$ in terms of the coefficients $\{a_1, a_2\}$. This allows us to explicitly derive $\hat S_{k}(f)$, and we have
\be
\hat S_{k}(f) = 
\left(
\begin{array}{cc}
r_k(f) & t_k(f) \\
t_k(-f) & r_k(f)
\end{array}
\right) = \hat S_{k}^{\rm T}(-f) \, ,
\label{S_2_terminals}
\ee
with
\begin{widetext}
\be
r_k(f) = \frac{6 \cos 2\pi kR + 2 \cos 2(\pi-\vartheta)kR - 8 \cos 2\pi R f}{-10\cos 2\pi kR + 2 \cos 2(\pi-\vartheta) kR + 8 \cos 2\pi R f + i \, 8 \sin 2\pi kR} \, = r_k(-f)
\label{r_k_2_terminals}
\ee
and
\be
t_k(f) = \frac{i \, 8\left[ \sin(2\pi-\vartheta) kR + {\rm e}^{-i2\pi Rf}\sin \vartheta kR\right] {\rm e}^{i\vartheta R f}}{-10\cos 2\pi kR + 2 \cos 2(\pi-\vartheta) kR + 8 \cos 2\pi R f + i \, 8 \sin 2\pi kR} \, .
\label{t_k_2_terminals}
\ee
\end{widetext}
It is worth noting that the time-reversal symmetry~\eqref{S-symm} satisfied by the scattering matrix can be readily checked on Eqs.~\eqref{S_2_terminals}-\eqref{t_k_2_terminals}.

\begin{figure}[h]
\begin{center}
\includegraphics[scale=0.55]{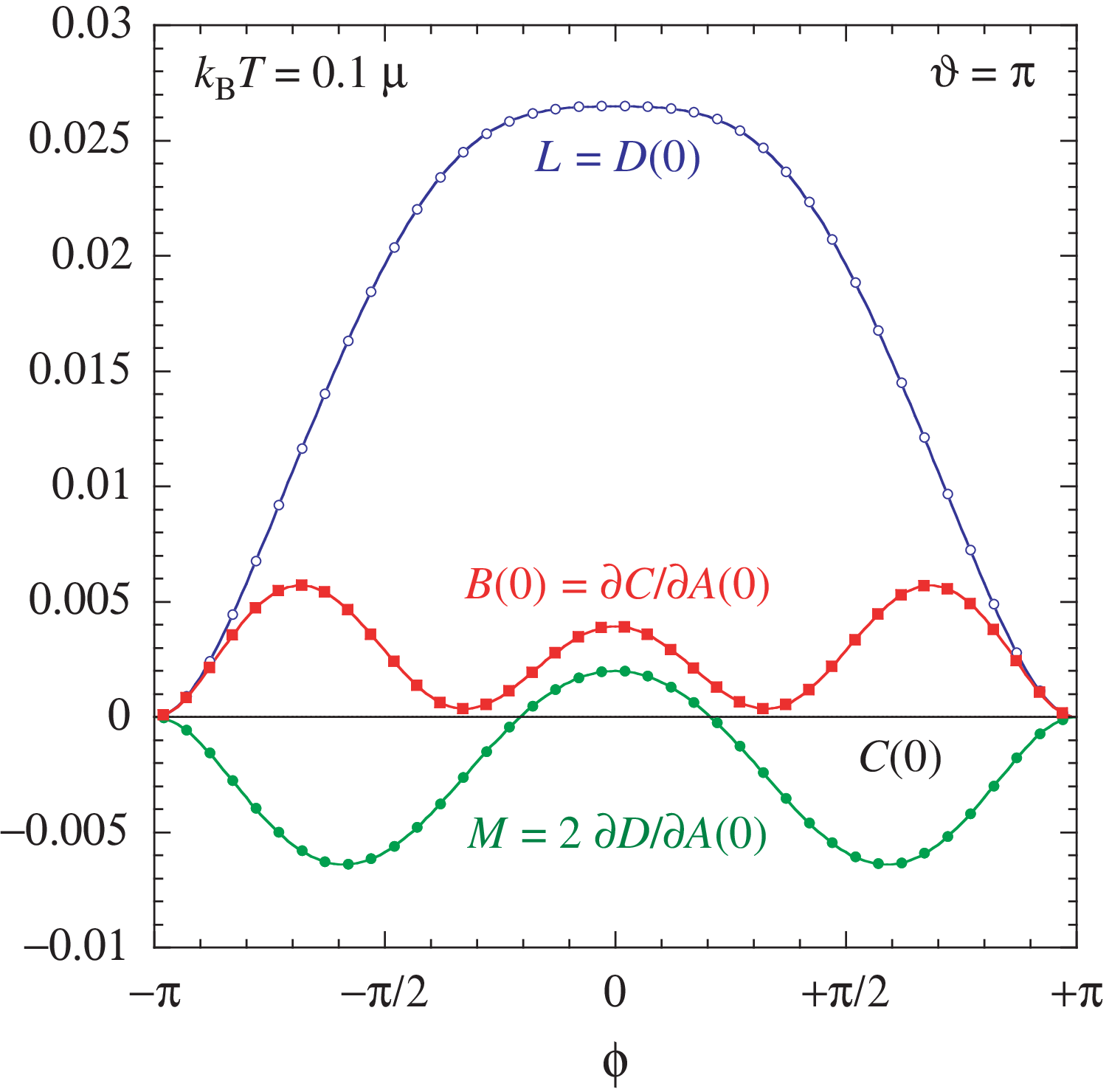}
\end{center}
\caption{Aharonov-Bohm ring with 2 terminals and $\vartheta=\pi$: Tests of the relations~\eqref{FDT}, \eqref{M-D-C-iii}, and \eqref{B0-C1}, together with the vanishing of the third cumulant~\eqref{C} at equilibrium for $k_{\rm B}T=0.1 \mu$, versus the dimensionless magnetic flux~\eqref{rescaled_mag_flux}.  The lines depict the left-hand side of the shown relations and the symbols their right-hand side.} 
\label{fig3}
\end{figure}

\begin{figure}[h]
\begin{center}
\includegraphics[scale=0.55]{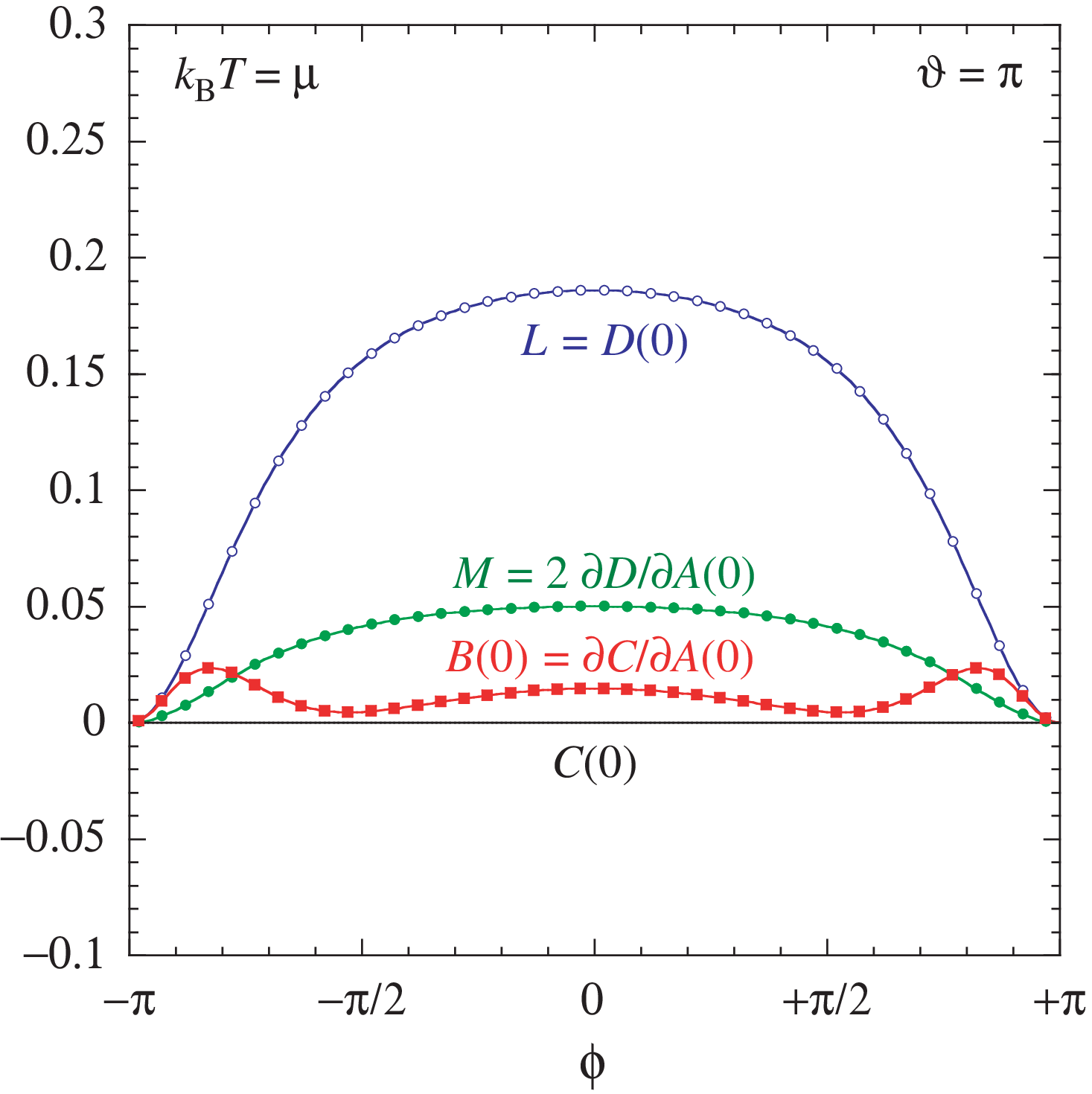}
\end{center}
\caption{Aharonov-Bohm ring with 2 terminals and $\vartheta=\pi$: Tests of the relations~\eqref{FDT}, \eqref{M-D-C-iii}, and \eqref{B0-C1}, together with the vanishing of the third cumulant~\eqref{C} at equilibrium for $k_{\rm B}T=\mu$, versus the dimensionless magnetic flux~\eqref{rescaled_mag_flux}. The lines depict the left-hand side of the shown relations and the symbols their right-hand side.} 
\label{fig4}
\end{figure}

\begin{figure}[h]
\begin{center}
\includegraphics[scale=0.55]{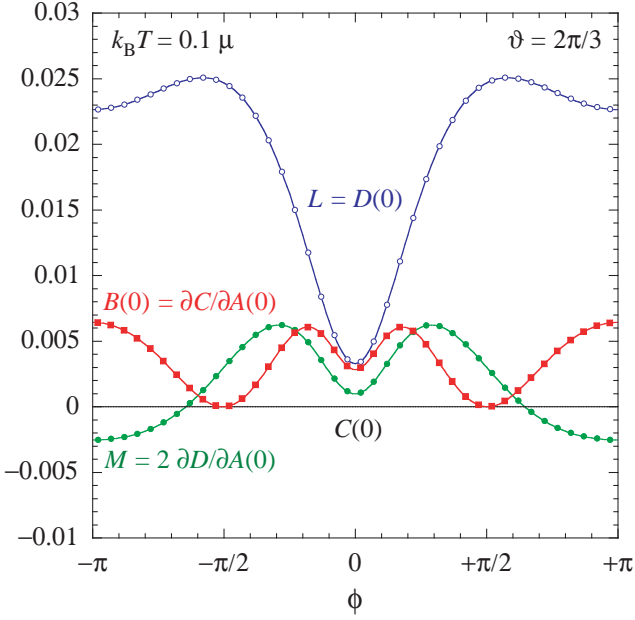}
\end{center}
\caption{Aharonov-Bohm ring with 2 terminals and $\vartheta=2\pi/3$: Tests of the relations~\eqref{FDT}, \eqref{M-D-C-iii}, and \eqref{B0-C1}, together with the vanishing of the third cumulant~\eqref{C} at equilibrium for $k_{\rm B}T=0.1 \mu$, versus the dimensionless magnetic flux~\eqref{rescaled_mag_flux}. The lines depict the left-hand side of the shown relations and the symbols their right-hand side.} 
\label{fig5}
\end{figure}

We now substitute the expressions~\eqref{S_2_terminals}-\eqref{t_k_2_terminals} into the derivatives~\eqref{R1}-\eqref{R3} in order to numerically evaluate the integrals~\eqref{Q1}-\eqref{Q3}, which yields the cumulants~\eqref{av_J}-\eqref{B}. This is done for a wide range of values of the dimensionless magnetic flux $\phi$ between $-\pi$ and $\pi$, allowing for a detailed study of the behavior of the cumulants with respect to the magnetic field [since the latter is related to $\phi$ through Eq.~\eqref{rescaled_mag_flux}]. We are thus able to illustrate some of the relations discussed in Sec.~\ref{transport_prop_sec}. We repeat this analysis for different values of the temperature or of the angle $\vartheta$ between the terminals.

We begin in Fig.~\ref{fig3} by taking an angle $\vartheta = \pi$ between the two terminals and a temperature such that $k_{\rm B}T=0.1\mu$. We first consider the relation~\eqref{FDT} between the (unique) diffusivity $D_{11}$ at equilibrium and the first response coefficient $L_{1,1}$, which merely reads here, in view of the Onsager-Casimir relation~\eqref{Casimir},
\be
D_{11}({\boldsymbol{0}};\boldsymbol{B}) = L_{1,1}(\boldsymbol{B}) \, .
\label{FDT_2_term}
\ee
The line $L$ in Fig.~\ref{fig3} plots the coefficient $L_{1,1}$, while the open circles correspond to the diffusivity $D_{11}$ in Eq.~\eqref{FDT_2_term}. The latter relation is thus perfectly confirmed by the numerical results. In addition, the symmetry of the curve $L=D(0)$ with respect to the axis $\phi=0$ clearly illustrates the fact that both $D_{11}({\boldsymbol{0}};\boldsymbol{B})$ and $L_{1,1}(\boldsymbol{B})$ are even functions of the magnetic field. Figure~\ref{fig3} also illustrates higher-order relations valid beyond the linear response regime, namely relations~\eqref{M-D-C-iii} and~\eqref{B0-C1}, as well as the vanishing of the (unique) third cumulant $C_{111}$ of this circuit. The latter property is indeed clearly demonstrated by the straight horizontal line. This readily simplifies the relation~\eqref{M-D-C-iii}, which merely reads here
\be
M_{1,11}(\boldsymbol{B}) = 2\, \frac{\partial D_{11}}{\partial A_{1}}({\boldsymbol{0}};\,\boldsymbol{B}) \, .
\label{M_D_A}
\ee
The left-hand side of this relation corresponds to the line $M$ and the right-hand side to the filled circles $2\partial D/\partial A(0)$ in Fig.~\ref{fig3}, which perfectly match each other. Hence, this confirms the validity of Eq.~\eqref{M_D_A}. Finally, the relation~\eqref{B0-C1} is illustrated by the line $B(0)$ and the filled squares $\partial C/\partial A(0)$, respectively depicting its left- and right-hand sides, and perfectly matching each other. This demonstrates in particular a symmetry property satisfied by the response $\partial C/\partial A(0)$ of the magnetic asymmetry $C_{111}$ to the (unique) affinity $A_1$ in the context of an Aharonov-Bohm ring connected to two terminals. Indeed, substituting the expression~\eqref{S_2_terminals} of the scattering matrix into Eq.~\eqref{Q3} with $\alpha=\beta=\gamma=1$, differentiating with respect to the affinity $A_1$, and setting $\boldsymbol{A}=\boldsymbol{0}$ in the resulting expression, we find that
\be
\frac{\partial C_{111}}{\partial A_{1}}(\boldsymbol{0};\boldsymbol{B}) = \frac{\partial C_{111}}{\partial A_{1}}(\boldsymbol{0};-\boldsymbol{B}) \, ,
\label{C_A_even}
\ee
thus giving $B(0)=B_{1111}({\boldsymbol{0}};\boldsymbol{B})$ according to Eq.~\eqref{B0-C1}.

We then repeat the exact same analysis in Fig.~\ref{fig4} and Fig.~\ref{fig5} for values $(\vartheta = \pi, k_{\rm B}T=\mu)$, and $(\vartheta = 2 \pi / 3, k_{\rm B}T=0.1\mu)$, respectively. Here again, the numerical results fully confirm the predictions drawn from the multivariate fluctuation relation~\eqref{sym-Q-magn}, i.e., from microreversibility in the presence of an external magnetic field.


\subsection{Aharonov-Bohm ring with 3 terminals}\label{AB_3_term_subsec}

\begin{figure}[t]
\begin{center}
\includegraphics[scale=0.35]{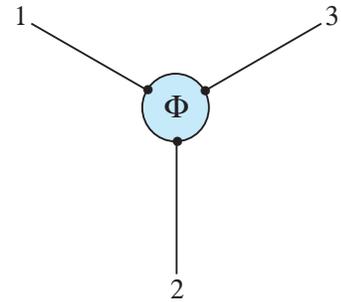}
\end{center}
\caption{Schematic representation of an Aharonov-Bohm ring with 3 terminals separated by angles $2 \pi /3$.} 
\label{fig6}
\end{figure}

The previous results are extended to an Aharonov-Bohm ring with three terminals separated by angles $2\pi/3$. That is, the angles $\theta_j$, $j = 1, 2, 3$, are here given by $\theta_1=0$, $\theta_2=2 \pi /3$, and $\theta_3=4 \pi /3$, as depicted in Fig.~\ref{fig6}.

The scattering matrix is here given by a $3 \times 3$ matrix. We then follow the exact same strategy as above, i.e., we solve the boundary conditions~\eqref{bc1}-\eqref{bc2} for $j=1,2,3$ and express the coefficients $\{b_1, b_2, b_3\}$ in terms of the coefficients $\{a_1, a_2, a_3\}$. This readily yields the scattering matrix $\hat S_{k}(f)$, and
\be
\hat S_{k}(f) = 
\left(
\begin{array}{ccc}
r_k(f) & t_k(f) & t_k(-f) \\
t_k(-f) & r_k(f) & t_k(f) \\
t_k(f) & t_k(-f) & r_k(f)
\end{array}
\right) = \hat S_{k}^{\rm T}(-f) \, ,
\label{S_3_terminals}
\ee
with
\begin{widetext}
\be
r_k(f) = \frac{6 \cos 2\pi kR + 2 \cos \frac{2\pi kR}{3} - 8 \cos 2\pi R f - i \, 3 \sin 2\pi kR - i \, 3 \sin\frac{2\pi kR}{3}}{-14\cos 2\pi kR +6 \cos \frac{2\pi kR}{3} + 8 \cos 2\pi R f + i \, 13 \sin 2\pi kR - i \, 3 \sin\frac{2\pi kR}{3}}
\, = r_k(-f)
\label{r_k_3_terminals}
\ee
and
\be
t_k(f) = \frac{i\, 8 \sin\frac{2\pi kR}{3} \left(2\cos \frac{2\pi kR}{3} - i \, \sin \frac{2\pi kR}{3} + {\rm e}^{-i2\pi Rf}\right)  {\rm e}^{i \frac{2\pi}{3} Rf}}{-14\cos 2\pi kR +6 \cos \frac{2\pi kR}{3} + 8 \cos 2\pi R f + i \, 13 \sin 2\pi kR - i \, 3 \sin\frac{2\pi kR}{3}}  \, ,
\label{t_k_3_terminals}
\ee
\end{widetext}
where the time-reversal symmetry~\eqref{S-symm} is again evident.  Similarly to Subsec.~\ref{AB_2_term_subsec}, we substitute the expressions~\eqref{S_3_terminals}-\eqref{t_k_3_terminals} into the derivatives~\eqref{R1}-\eqref{R3}. We then numerically evaluate the integrals~\eqref{Q1}-\eqref{Q3} in order to obtain the cumulants~\eqref{av_J}-\eqref{B} for a range of values of $\phi$ between $-\pi$ and $\pi$. This again allows us to test various relations discussed in Sec.~\ref{transport_prop_sec}. All the numerical calculations presented in this subsection have been done for $k_{\rm B}T=0.1\mu$.

\begin{figure}[h]
\begin{center}
\includegraphics[scale=0.55]{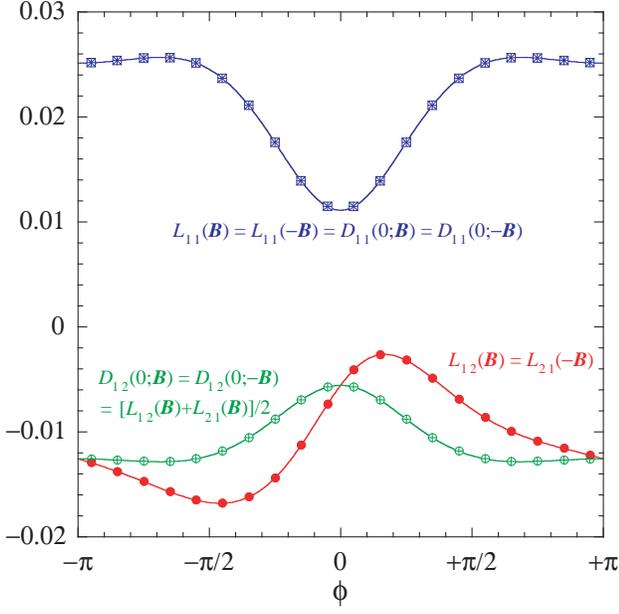}
\end{center}
\caption{Aharonov-Bohm ring with 3 terminals: Tests of the symmetries~\eqref{Casimir} and~\eqref{FDT} concerning the linear transport properties for $k_{\rm B}T=0.1 \mu$, versus the dimensionless magnetic flux~\eqref{rescaled_mag_flux}. The lines depict the left-hand side of the shown relations and the symbols the other members of the relations. $L_{1,1}(-\boldsymbol{B})$ is depicted by crosses,  $D_{11}({\boldsymbol{0}};\boldsymbol{B})$ by open squares, $D_{11}({\boldsymbol{0}};-\boldsymbol{B})$ and $D_{12}({\boldsymbol{0}};-\boldsymbol{B})$ by pluses, $L_{2,1}(-\boldsymbol{B})$ by filled dots, and the right-hand side of Eq.~\eqref{FDT} for $\alpha,\beta=1,2$ by open circles.} 
\label{fig7}
\end{figure}

\begin{figure}[h]
\begin{center}
\includegraphics[scale=0.55]{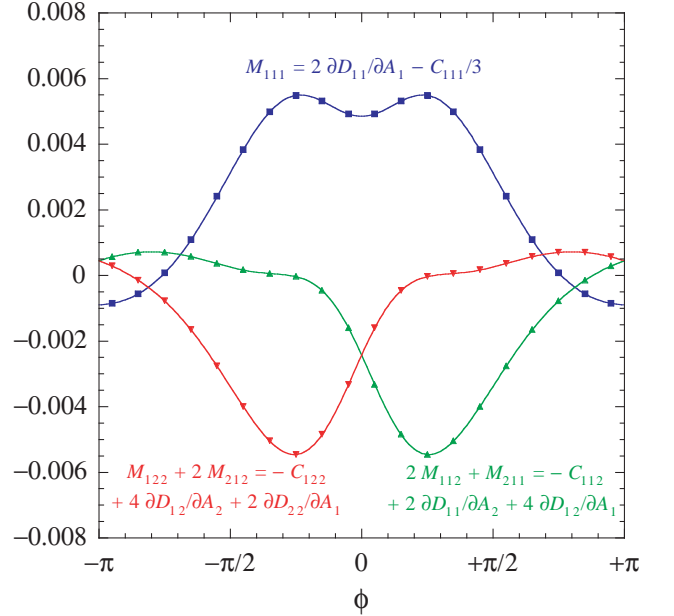}
\end{center}
\caption{Aharonov-Bohm ring with 3 terminals: Tests of the relation~\eqref{M-D1-C0} between nonlinear transport properties for $k_{\rm B}T=0.1 \mu$, versus the dimensionless magnetic flux~\eqref{rescaled_mag_flux}.  The lines depict the left-hand side of the shown relations and the symbols their right-hand side.} 
\label{fig8}
\end{figure}

\begin{figure}[h]
\begin{center}
\includegraphics[scale=0.55]{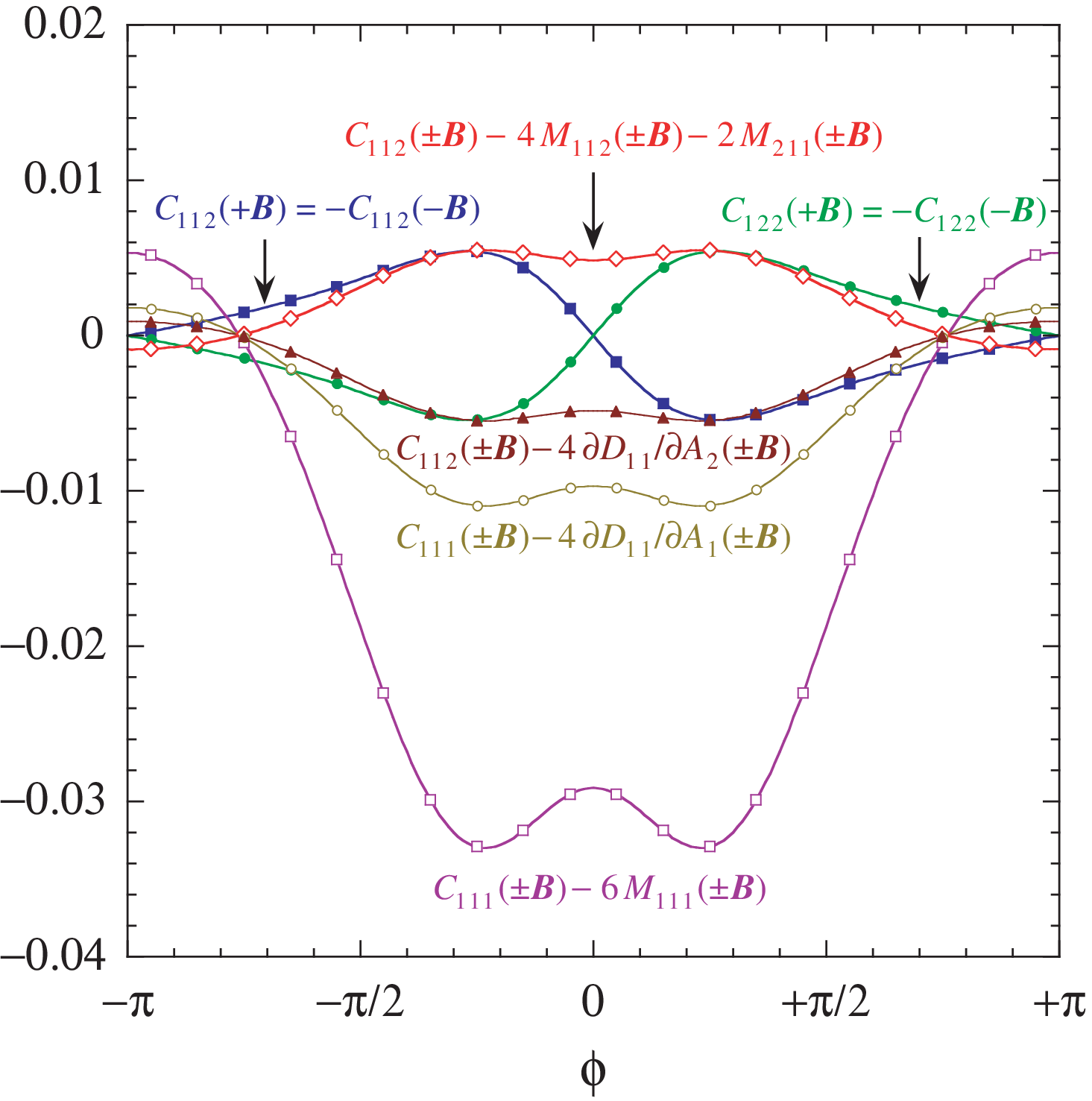}
\end{center}
\caption{Aharonov-Bohm ring with 3 terminals: Tests of the symmetries~\eqref{C0-C0}, \eqref{C0-D1}, and~\eqref{C-M-odd} concerning the nonlinear transport properties for $k_{\rm B}T=0.1 \mu$, versus the dimensionless magnetic flux~\eqref{rescaled_mag_flux}.  The lines depict the relations for $\boldsymbol{B}$ and the symbols those for $-\boldsymbol{B}$.} 
\label{fig9}
\end{figure}

\begin{figure}[h]
\begin{center}
\includegraphics[scale=0.55]{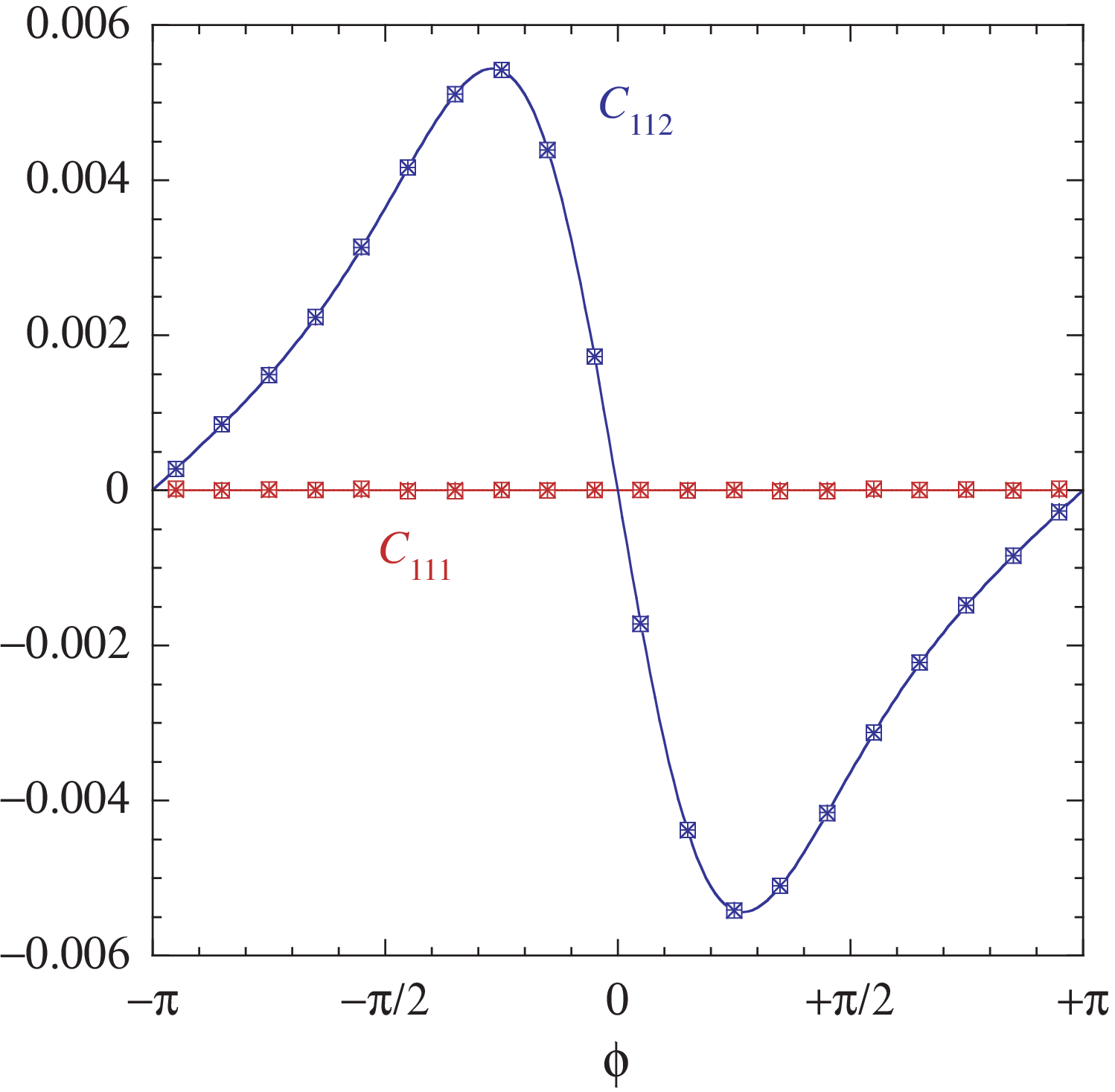}
\end{center}
\caption{Aharonov-Bohm ring with 3 terminals: Tests of the relations~\eqref{C0-D1}, \eqref{M-D1-C0}, and~\eqref{C-M-odd} between nonlinear transport properties for $k_{\rm B}T=0.1 \mu$, versus the dimensionless magnetic flux~\eqref{rescaled_mag_flux}.  The lines depict the third cumulant~\eqref{C}. The right-hand sides of the aforementioned relations are respectively depicted as pluses, open squares, and crosses.} 
\label{fig10}
\end{figure}

\begin{figure}[b]
\begin{center}
\includegraphics[scale=0.55]{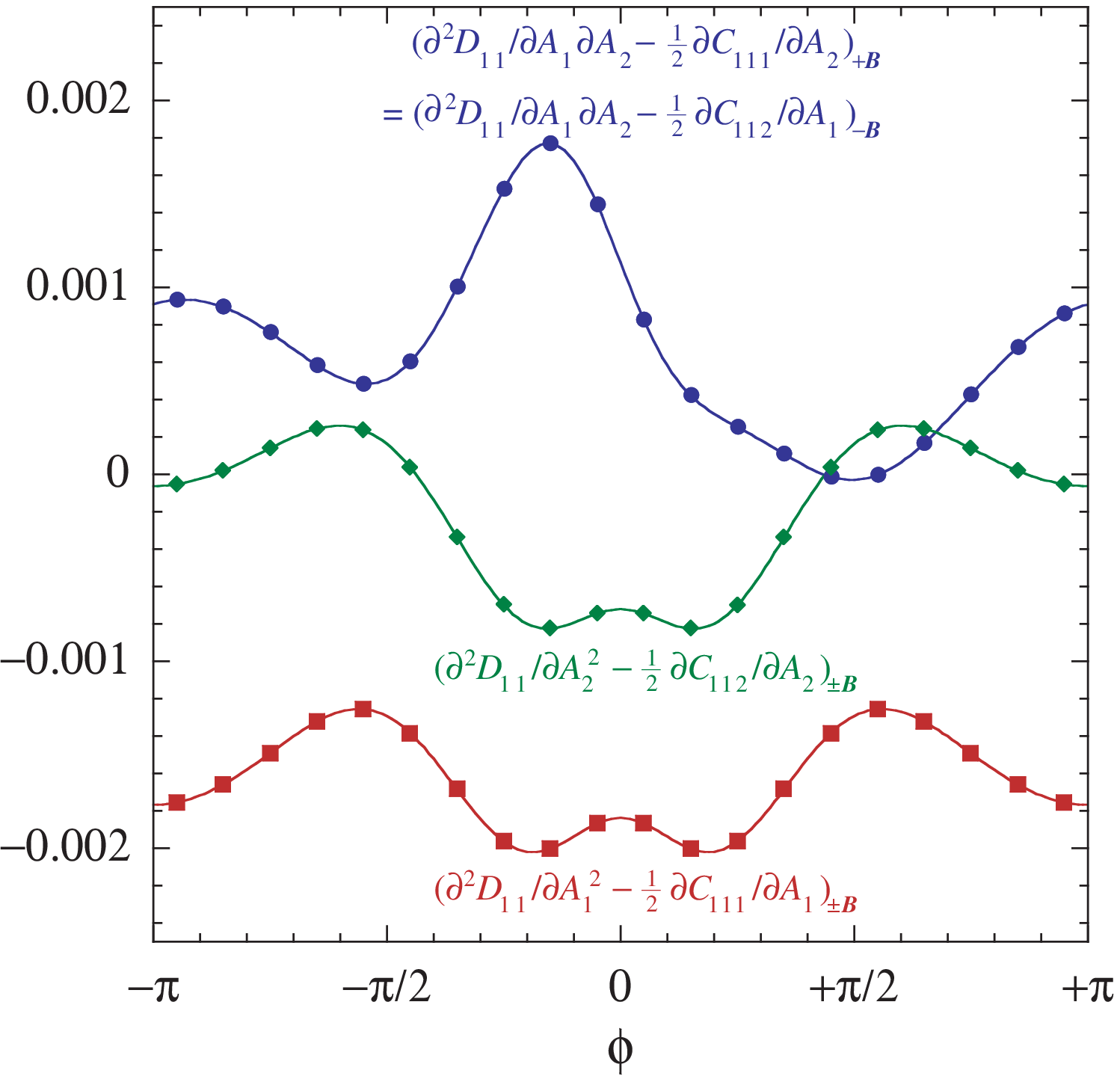}
\end{center}
\caption{Aharonov-Bohm ring with 3 terminals: Tests of the relation~\eqref{D2-C1} between nonlinear transport properties for $k_{\rm B}T=0.1 \mu$, versus the dimensionless magnetic flux~\eqref{rescaled_mag_flux}. The lines depict the relations for $\boldsymbol{B}$ and the symbols those for $-\boldsymbol{B}$.} 
\label{fig11}
\end{figure}

\begin{figure}[t]
\begin{center}
\includegraphics[scale=0.55]{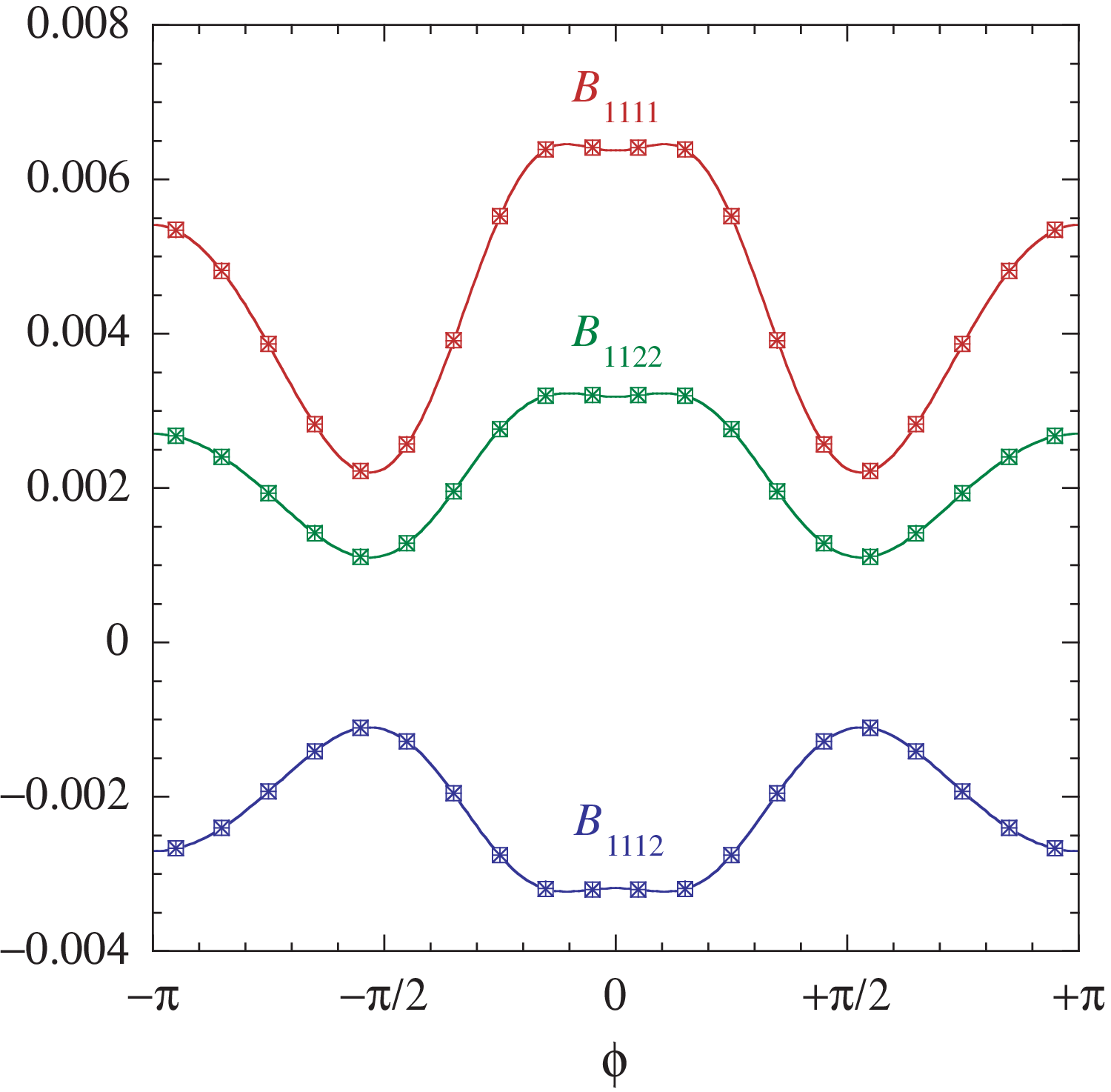}
\end{center}
\caption{Aharonov-Bohm ring with 3 terminals: Tests of the relations~\eqref{B0-C1}, \eqref{N-N}, and~\eqref{N-D2-C1-B0} between nonlinear transport properties for $k_{\rm B}T=0.1 \mu$, versus the dimensionless magnetic flux~\eqref{rescaled_mag_flux}. The lines depict the fourth cumulants~\eqref{B}. The right-hand sides of the aforementioned relations are respectively depicted as open squares, crosses, and pluses.} 
\label{fig12}
\end{figure}

We first illustrate in Fig.~\ref{fig7} the symmetries~\eqref{Casimir} and~\eqref{FDT}. The solid  lines respectively show the behavior of $L_{1,1}(\boldsymbol{B})$, $L_{1,2}(\boldsymbol{B})$, and $D_{12}({\boldsymbol{0}};\boldsymbol{B})$ with respect to the magnetic field [embedded in the dimensionless magnetic flux $\phi$ according to Eq.~\eqref{rescaled_mag_flux}], while the symbols depict the other members of the illustrated relations. It is clearly seen that the symmetry $L_{1,1}(\boldsymbol{B}) = L_{1,1}(-\boldsymbol{B})$, i.e., precisely Eq.~\eqref{Casimir} for $\alpha = \beta = 1$, is indeed satisfied. Incidentally, the relation~\eqref{FDT} for $\alpha = \beta = 1$ actually reads as Eq.~\eqref{FDT_2_term}. The diffusivity $D_{11}({\boldsymbol{0}};\boldsymbol{B})$ is depicted by open squares, which confirms the symmetry~\eqref{FDT_2_term}. Furthermore, the diffusivity $D_{11}({\boldsymbol{0}};-\boldsymbol{B})$ for a reversed magnetic field is depicted by pluses. The even behavior of $D_{11}$ with respect to the magnetic field is then obvious from the plot. Moreover, the filled circles show $L_{2,1}(-\boldsymbol{B})$ and they are superimposed on the solid line representing the first response coefficient $L_{1,2}(\boldsymbol{B})$. This again confirms the Onsager-Casimir reciprocity relation~\eqref{Casimir} for $\alpha, \beta = 1,2$. For these values of $\alpha$ and $\beta$, the relation~\eqref{FDT} now reads, also with Eq.~\eqref{Casimir},
\be
D_{1 2}({\boldsymbol{0}};\boldsymbol{B}) = \frac{1}{2} \left[ L_{1,2}(\boldsymbol{B}) + L_{2,1}(\boldsymbol{B}) \right] \, .
\label{FDT_3_term_bis}
\ee
This relation is fully confirmed since the open circles [representing the right-hand side of Eq.~\eqref{FDT_3_term_bis}] perfectly match the solid line showing the diffusivity $D_{1 2}({\boldsymbol{0}};\boldsymbol{B})$. Finally, the even behavior of the latter with respect to the magnetic field is illustrated by plotting $D_{1 2}({\boldsymbol{0}};-\boldsymbol{B})$ with pluses.

We then illustrate in Fig.~\ref{fig8} the relation~\eqref{M-D1-C0} between nonlinear transport properties for different values of the indices $\alpha$, $\beta$, and $\gamma$. Taking first $\alpha = \beta = \gamma = 1$ into Eq.~\eqref{M-D1-C0} yields
\be
M_{1,11} (\boldsymbol{B}) = 2\, \frac{\partial D_{11}}{\partial A_{1}} ({\boldsymbol{0}};\boldsymbol{B}) - \frac{1}{3}\, C_{111}({\boldsymbol{0}};\boldsymbol{B}) \; .
\label{M_D1_C0_3_term}
\ee
The solid line showing the second response coefficient $M_{1,11} (\boldsymbol{B})$ coincides with the filled squares depicting the right-hand side of Eq.~\eqref{M_D1_C0_3_term}. We can thus readily see that Eq.~\eqref{M_D1_C0_3_term} is indeed satisfied. In addition, we test the relation~\eqref{M-D1-C0} for $\alpha = \beta = 1$ and $\gamma = 2$, namely
\begin{multline}
2 M_{1,12} (\boldsymbol{B}) + M_{2,11} (\boldsymbol{B}) = 2\, \frac{\partial D_{11}}{\partial A_{2}} ({\boldsymbol{0}};\boldsymbol{B}) \\
+ 4\, \frac{\partial D_{12}}{\partial A_{1}} ({\boldsymbol{0}};\boldsymbol{B}) - C_{112}({\boldsymbol{0}};\boldsymbol{B}) \, ,
\label{M_D1_C0_3_term_bis}
\end{multline}
and for $\alpha = 1$ and $\beta = \gamma = 2$, namely
\begin{multline}
M_{1,22} (\boldsymbol{B}) + 2 M_{2,12} (\boldsymbol{B}) = 4\, \frac{\partial D_{12}}{\partial A_{2}} ({\boldsymbol{0}};\boldsymbol{B}) \\
+ 2\, \frac{\partial D_{22}}{\partial A_{1}} ({\boldsymbol{0}};\boldsymbol{B}) - C_{122}({\boldsymbol{0}};\boldsymbol{B}) \, ,
\label{M_D1_C0_3_term_ter}
\end{multline}
since the quantities $M_{\alpha , \beta \gamma}$ and $D_{\beta \gamma}$ are invariant under a permutation of the indices $\beta$ and $\gamma$. The left-hand sides of Eqs.~\eqref{M_D1_C0_3_term_bis} and~\eqref{M_D1_C0_3_term_ter} are shown by the solid lines with superimposed up and down triangles depicting their right-hand sides. The two above relations are thus again satisfied.

Figure~\ref{fig9} then tests various properties of the third cumulants $C_{\alpha \beta \gamma}({\boldsymbol{0}};\boldsymbol{B})$. It first illustrates the fact that they are odd with respect to the magnetic field, which corresponds to the property~\eqref{C0-C0}. The lines with filled squares and circles show that the relation~\eqref{C0-C0} holds for $(\alpha = \beta = 1,\gamma = 2)$ and~$(\alpha = 1,\beta = \gamma = 2)$. Moreover, the solid lines with open circles and triangles confirm for $(\alpha = \beta = \gamma = 1)$ and $(\alpha = \beta = 1,\gamma = 2)$ the validity of
\begin{multline}
C_{\alpha\beta\gamma}({\boldsymbol{0}};\boldsymbol{B}) - 4\, 
\frac{\partial D_{\alpha\beta}}{\partial A_{\gamma}}({\boldsymbol{0}};\boldsymbol{B}) \\
= C_{\alpha\beta\gamma}({\boldsymbol{0}};-\boldsymbol{B}) - 4\, 
\frac{\partial D_{\alpha\beta}}{\partial A_{\gamma}}({\boldsymbol{0}};-\boldsymbol{B}) \, ,
\end{multline}
which is deduced by combining Eq.~\eqref{C0-D1} with Eq.~\eqref{C0-C0}.  This clearly shows that these quantities are even with respect to the magnetic field.  Similarly, combining Eq.~\eqref{C-M-odd} with Eq.~\eqref{C0-C0} gives
\begin{multline}
C_{\alpha\beta\gamma}({\boldsymbol{0}};\boldsymbol{B}) - \left(M_{\alpha,\beta\gamma}+M_{\beta,\gamma\alpha}+M_{\gamma,\alpha\beta}\right)_{\boldsymbol{B}} \\
= C_{\alpha\beta\gamma}({\boldsymbol{0}};-\boldsymbol{B}) -\left(M_{\alpha,\beta\gamma}+M_{\beta,\gamma\alpha}+M_{\gamma,\alpha\beta}\right)_{-\boldsymbol{B}} \; ,
\label{C-M-odd-bis}
\end{multline}
which are confirmed by the solid lines (showing left-hand sides) with the open squares and open diamonds (depicting right-hand sides), respectively for $(\alpha = \beta = \gamma = 1)$ and~$(\alpha = \beta = 1,\gamma = 2)$. The two quantities are again seen to be the same.

Figure~\ref{fig10} further tests some of the properties of the third cumulants $C_{111}({\boldsymbol{0}};\boldsymbol{B})$ and $C_{112}({\boldsymbol{0}};\boldsymbol{B})$. First, we note that the cumulant $C_{111}$ vanishes identically, as is expected and confirmed by the straight horizontal line. The other line then shows the cumulant $C_{112}$. These cumulants can be alternatively expressed in terms of the quantities $M$ and/or $\partial D / \partial A$ according to expressions~\eqref{C0-D1}, \eqref{M-D1-C0}, or~\eqref{C-M-odd} for $(\alpha = \beta =\gamma = 1)$ or $(\alpha = \beta = 1,\gamma = 2)$. These relations are fully confirmed, as we note that their right-hand sides, respectively depicted by pluses, open squares and crosses, perfectly match the corresponding solid line. Finally, as shown in Sec.~\ref{transport_prop_sec}, we have that $C_{111}=C_{222}=0$ at equilibrium.

We then verify in Fig.~\ref{fig11} the relation~\eqref{D2-C1} between the second responses $\partial^2 D / \partial A^2$ of the diffusivities and the first responses $\partial C / \partial A$ of the third cumulants. The lines with filled squares, circles, and diamonds test the relations~\eqref{D2-C1} respectively for $(\alpha = \beta = \gamma = \delta = 1)$, $(\alpha = \beta = \gamma = 1,\delta = 2)$, and $(\alpha = \beta = 1,\gamma = \delta = 2)$, which clearly demonstrates their validity.

Finally, Fig.~\ref{fig12} tests the relations~\eqref{B0-C1}, \eqref{N-N}, and~\eqref{N-D2-C1-B0}, which alternatively express the fourth cumulants $B$ in terms of the quantities $\partial C / \partial A$, $N$, and/or $\partial^2 D / \partial A^2$. The solid lines show the cumulants $B_{1111}$, $B_{1112}$ and $B_{1122}$. The right-hand sides of the corresponding expressions~\eqref{B0-C1}, \eqref{N-N}, and~\eqref{N-D2-C1-B0} of these cumulants are depicted by open squares, crosses and pluses, respectively, which are readily seen to perfectly match each of the corresponding solid lines. This demonstrates the validity of the expressions~\eqref{B0-C1}, \eqref{N-N}, and~\eqref{N-D2-C1-B0}.
 
\vskip 0.5 cm

\subsection{Aharonov-Bohm ring with 4 terminals}\label{AB_4_term_subsec}

\begin{figure}[h]
\begin{center}
\includegraphics[scale=0.35]{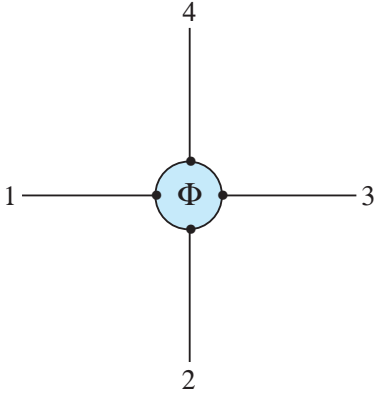}
\end{center}
\caption{Schematic representation of an Aharonov-Bohm ring with 4 terminals separated by angles $\pi / 2$.} 
\label{fig13}
\end{figure}

We conclude our analysis by considering an Aharonov-Bohm ring with four terminals forming right angles, as schematically represented in Fig.~\ref{fig13}. That is, the terminals are attached to the ring at the angles $\theta_1=0$, $\theta_2=\pi/2$, $\theta_3=\pi$, and $\theta_4=3\pi/2$. Here again, we solve the boundary conditions~\eqref{bc1}-\eqref{bc2} for $j=1,2,3,4$ and express the coefficients $\{b_1, b_2, b_3, b_4\}$ in terms of $\{a_1, a_2, a_3, a_4\}$ to get the $4 \times 4$ scattering matrix
\be
\hat S_{k}(f) = 
\left(
\begin{array}{cccc}
r_k(f) & s_k(f) & t_k(f) & s_k(-f) \\
s_k(-f) & r_k(f) & s_k(f)  & t_k(f) \\
t_k(-f) & s_k(-f) & r_k(f) & s_k(f) \\
s_k(f) & t_k(-f) & s_k(-f) & r_k(f)
\end{array}
\right) = \hat S_{k}^{\rm T}(-f) \, ,
\label{S_4_terminals}
\ee
with
\begin{widetext}
\be
r_k(f) = \frac{3-15 \cos 2\pi kR -4 \cos \pi kR + 16 \cos 2\pi Rf + i \, 12 \sin 2\pi kR + i \, 8 \sin \pi kR}{-5 +41 \cos 2\pi kR -20 \cos \pi kR -16 \cos 2\pi Rf -i \, 40 \sin 2 \pi kR + i \, 16 \sin \pi kR} \, = r_k(-f) \, ,
\label{r_k_4_terminals}
\ee
\be
t_k(f) = \frac{-16 \cos\pi Rf \left( 1 - \cos \pi kR + i \, 2 \sin \pi kR\right)}{-5 +41 \cos 2\pi kR -20 \cos \pi kR -16 \cos 2\pi Rf -i \, 40 \sin 2 \pi kR + i \, 16 \sin \pi kR} \, = t_k(-f) \, ,
\label{t_k_4_terminals}
\ee
and
\be
s_k(f) = \frac{-i \, 4  \sin\frac{\pi kR}{2} \left( 2 + 10 \cos \pi kR - i \, 8 \sin \pi kR + 4 \, {\rm e}^{-i 2\pi Rf}\right){\rm e}^{i \frac{\pi}{2} Rf}}{-5 +41 \cos 2\pi kR -20 \cos \pi kR -16 \cos 2\pi Rf -i \, 40 \sin 2 \pi kR + i \, 16 \sin \pi kR} \, ,
\label{s_k_4_terminals}
\ee
\end{widetext}
which indeed demonstrates the time-reversal symmetry~\eqref{S-symm}. Substituting again the expressions~\eqref{S_4_terminals}-\eqref{s_k_4_terminals} into the derivatives~\eqref{R1}-\eqref{R3} allows us to numerically evaluate the integrals~\eqref{Q1}-\eqref{Q3} and obtain the cumulants~\eqref{av_J}-\eqref{B} in a range of values of $\phi$ between $-\pi$ and $\pi$ [i.e., a range of values of $B$ in view of Eq.~\eqref{rescaled_mag_flux}]. We then extend the analysis performed in Subsec.~\ref{AB_3_term_subsec} in the case of an Aharonov-Bohm ring with three terminals. Here again, all the numerical results have been obtained for $k_{\rm B}T=0.1\mu$.

\begin{figure}[h]
\begin{center}
\includegraphics[scale=0.55]{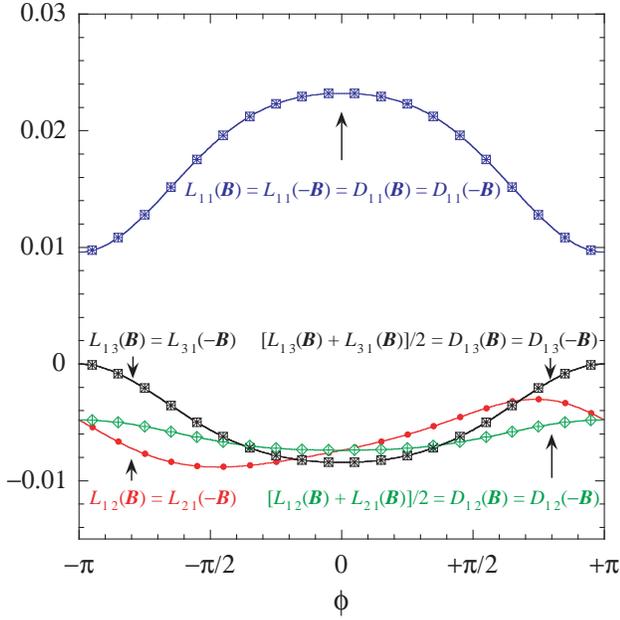}
\end{center}
\caption{Aharonov-Bohm ring with 4 terminals: Tests of the symmetries~\eqref{Casimir} and~\eqref{FDT} concerning the linear transport properties for $k_{\rm B}T=0.1 \mu$, versus the dimensionless magnetic flux~\eqref{rescaled_mag_flux}.  The lines depict the left-hand sides of the shown relations. $L_{11}(-\boldsymbol{B})$ and $L_{31}(-\boldsymbol{B})$ are depicted by open squares; $D_{11}({\boldsymbol{0}};\boldsymbol{B})$ and $D_{13}({\boldsymbol{0}};-\boldsymbol{B})$ by crosses; $D_{11}({\boldsymbol{0}};-\boldsymbol{B})$, $D_{13}({\boldsymbol{0}};\boldsymbol{B})$, and $D_{12}({\boldsymbol{0}};-\boldsymbol{B})$ by pluses; $L_{21}(-\boldsymbol{B})$ by dots; and $D_{12}(\boldsymbol{B})$ by open diamonds.} 
\label{fig14}
\end{figure}

\begin{figure}[h]
\begin{center}
\includegraphics[scale=0.55]{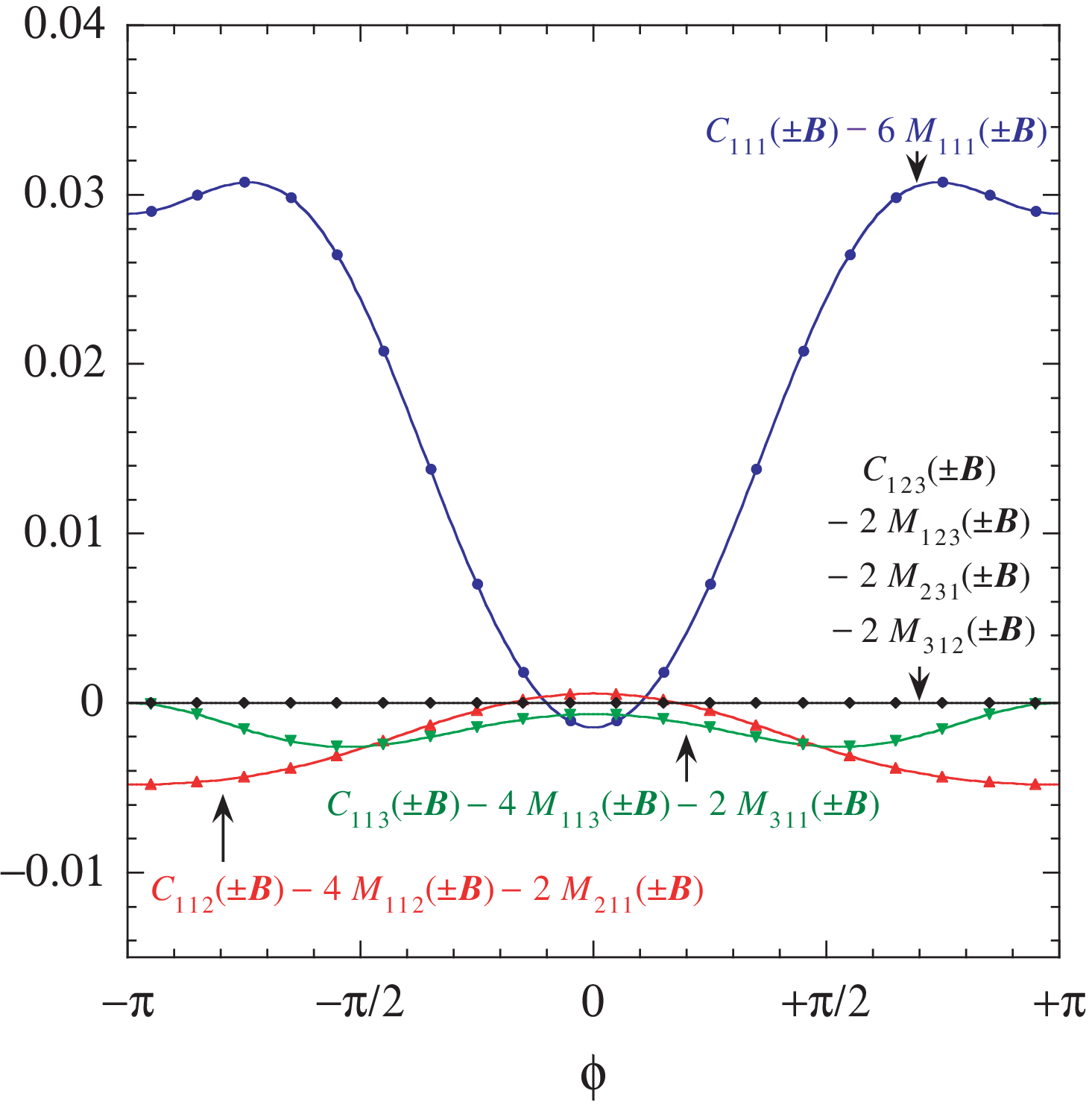}
\end{center}
\caption{Aharonov-Bohm ring with 4 terminals: Tests of the relation~\eqref{C-M-odd} between nonlinear transport properties for $k_{\rm B}T=0.1 \mu$, versus the dimensionless magnetic flux~\eqref{rescaled_mag_flux}. The lines depict the relations for $\boldsymbol{B}$ and the symbols those for $-\boldsymbol{B}$.} 
\label{fig15}
\end{figure}

\begin{figure}[t]
\begin{center}
\includegraphics[scale=0.55]{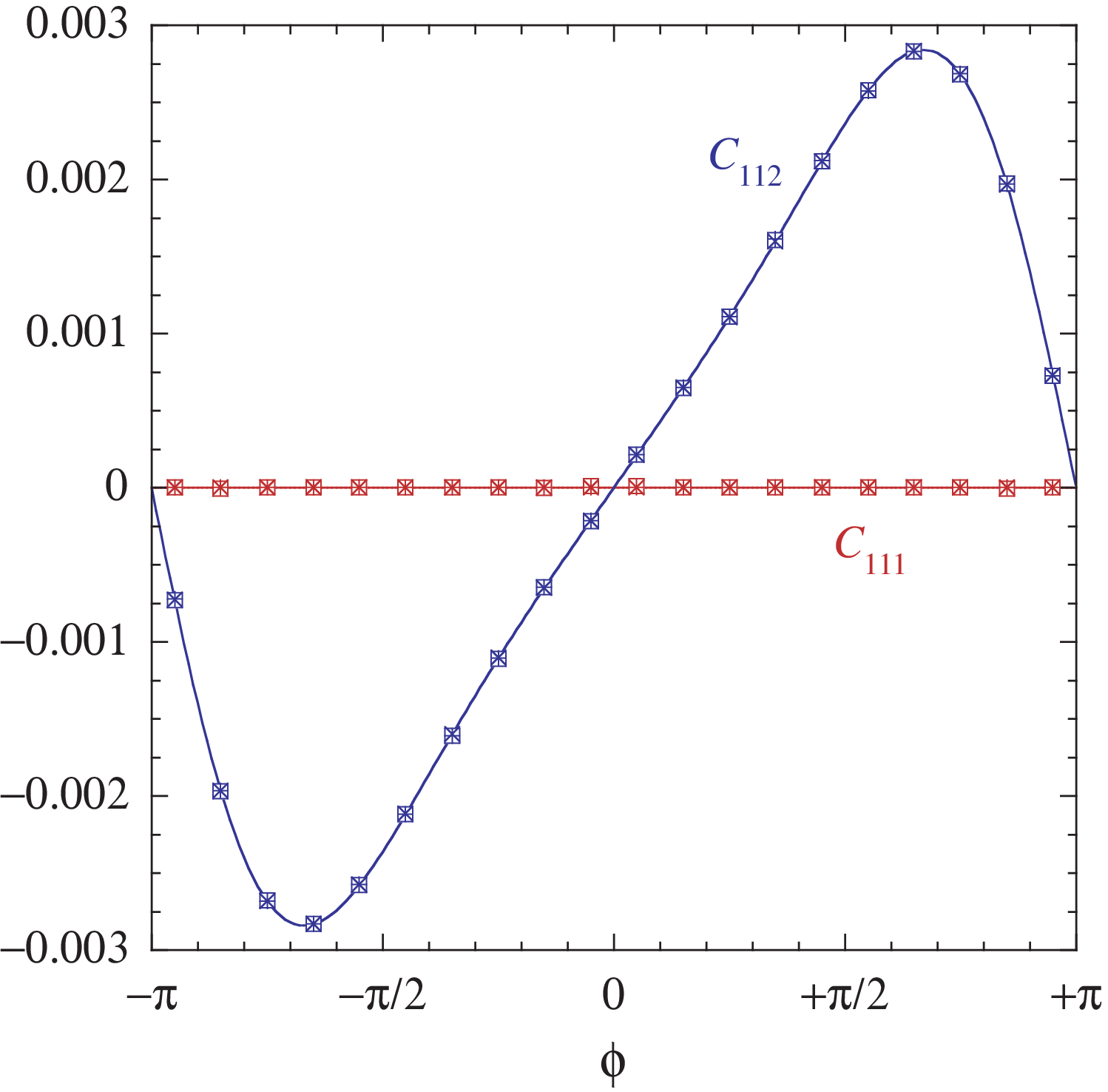}
\end{center}
\caption{Aharonov-Bohm ring with 4 terminals: Tests of the relations~\eqref{C0-D1}, \eqref{M-D1-C0}, and~\eqref{C-M-odd} between nonlinear transport properties for $k_{\rm B}T=0.1 \mu$, versus the dimensionless magnetic flux~\eqref{rescaled_mag_flux}.  The lines depict the third cumulants~\eqref{C}.  The right-hand sides of the aforementioned relations are respectively depicted as pluses, open squares, and crosses.} 
\label{fig16}
\end{figure}

\begin{figure}[t]
\begin{center}
\includegraphics[scale=0.55]{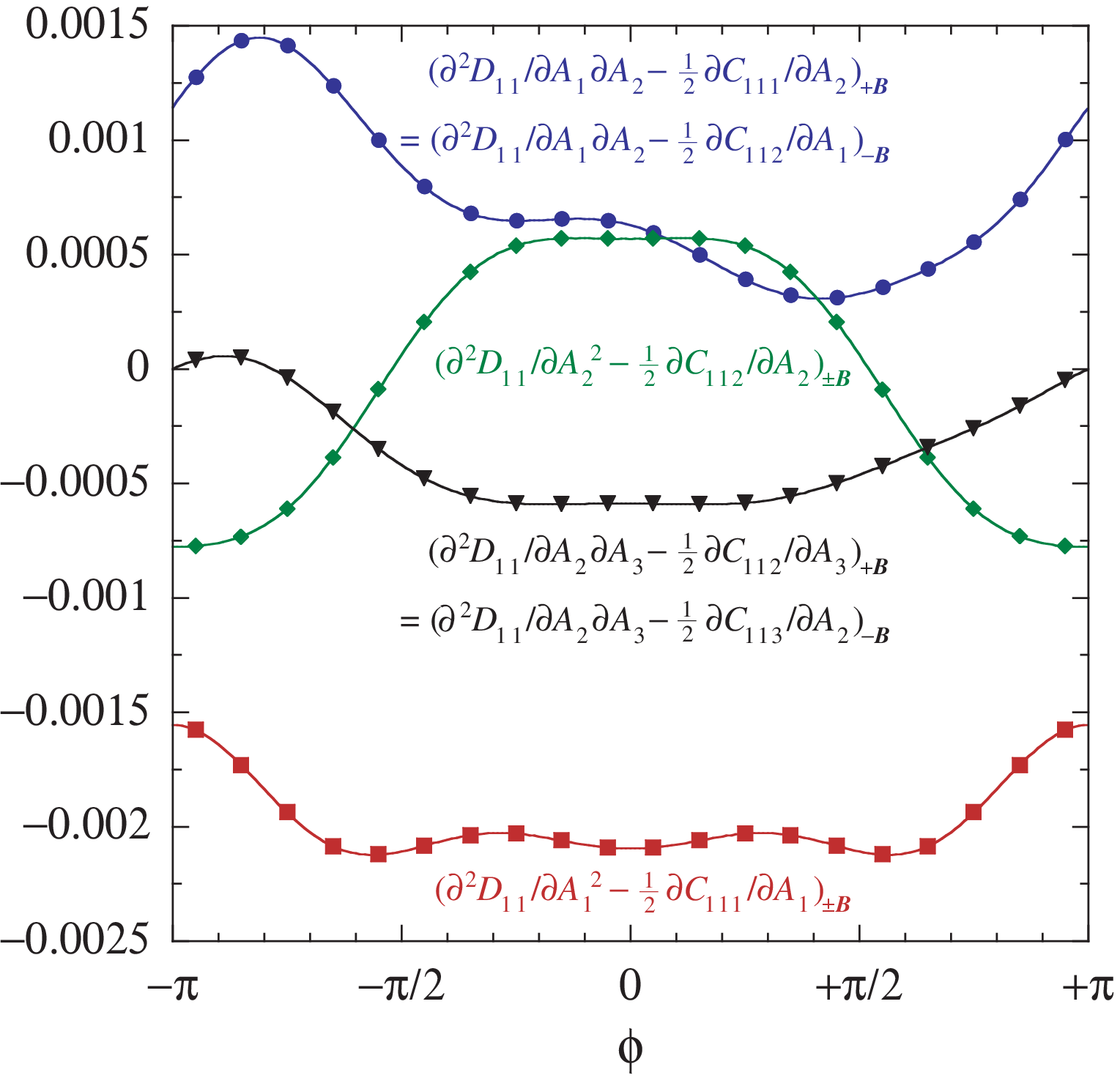}
\end{center}
\caption{Aharonov-Bohm ring with 4 terminals: Tests of the relation~\eqref{D2-C1} between nonlinear transport properties for $k_{\rm B}T=0.1 \mu$, versus the dimensionless magnetic flux~\eqref{rescaled_mag_flux}. The lines depict the relations for $\boldsymbol{B}$ and the symbols those for $-\boldsymbol{B}$.} 
\label{fig17}
\end{figure}

\begin{figure}[b]
\begin{center}
\includegraphics[scale=0.55]{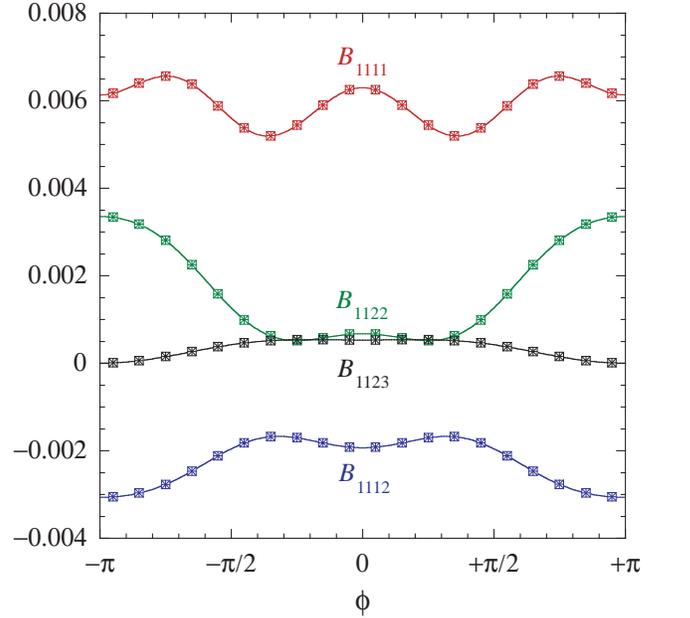}
\end{center}
\caption{Aharonov-Bohm ring with 4 terminals: Tests of the relations~\eqref{B0-C1}, \eqref{N-N}, and~\eqref{N-D2-C1-B0} between nonlinear transport properties for $k_{\rm B}T=0.1 \mu$, versus the dimensionless magnetic flux~\eqref{rescaled_mag_flux}.  The lines depict the fourth cumulants~\eqref{B}. The right-hand sides of the aforementioned relations are respectively depicted as open squares, crosses, and pluses.} 
\label{fig18}
\end{figure}

We begin in Fig.~\ref{fig14} with the symmetries~\eqref{Casimir} and~\eqref{FDT}. The solid line with the open squares represent the first response coefficients $L_{1,1}(\boldsymbol{B})$ and $L_{1,1}(-\boldsymbol{B})$, respectively. This readily demonstrates the validity of the Onsager-Casimir relation~\eqref{Casimir} for $\alpha = \beta = 1$, i.e., that $L_{1,1}$ is even with respect to the magnetic field. A direct consequence is that the relation~\eqref{FDT} for $\alpha = \beta = 1$ yields $D_{11}({\bf 0};\boldsymbol{B}) = L_{1,1}(\boldsymbol{B}) = D_{11}({\boldsymbol{0}};-\boldsymbol{B})$. This is readily confirmed by our numerical results, as the diffusivities $D_{11}({\boldsymbol{0}};\boldsymbol{B})$ and $D_{11}({\boldsymbol{0}};-\boldsymbol{B})$, depicted by crosses and pluses, respectively, indeed perfectly match the corresponding solid line.  The Onsager-Casimir relation~\eqref{Casimir} is then illustrated for $\alpha = 1$ and $\beta = 2$. The coefficients $L_{1,2}(\boldsymbol{B})$ and $L_{2,1}(-\boldsymbol{B})$ are represented by the solid line with filled circles and they overlap perfectly. Here again, the relation~\eqref{FDT} for $\alpha, \beta = 1,2$ can be written as Eq.~\eqref{FDT_3_term_bis}, which is confirmed since the open diamonds and the pluses, respectively depicting $D_{1 2}({\boldsymbol{0}};\boldsymbol{B})$ and $D_{1 2}({\boldsymbol{0}};-\boldsymbol{B})$, are superimposed on the line giving $[L_{1,2}(\boldsymbol{B})+L_{2,1}(\boldsymbol{B})]/2$.  Moreover, for $\alpha = 1$ and $\beta = 3$, the angle between the terminals is equal to $\theta_3-\theta_1=\pi$.  As a consequence, $L_{1,3}(\boldsymbol{B})=L_{3,1}(\boldsymbol{B})=L_{1,3}(-\boldsymbol{B})$ is even in the magnetic field, as can be explicitly checked. Indeed, substituting the expression~\eqref{S_4_terminals} of the scattering matrix into the expression~\eqref{Q1}, setting $\alpha=1$ and then $\alpha=3$, differentiating with respect to the affinity $A_3$ and then $A_1$, setting $\boldsymbol{A}=\boldsymbol{0}$ in the resulting expressions and exploiting the symmetry $t_k(f)=t_k(-f)$ [see Eq.~\eqref{t_k_4_terminals}] shows that $L_{1,3}(\boldsymbol{B})=L_{3,1}(\boldsymbol{B})$. This readily implies $L_{1,3}(\boldsymbol{B})=L_{1,3}(-\boldsymbol{B})$ in view of the Onsager-Casimir relation~\eqref{Casimir} for $\alpha=3$ and $\beta=1$. This is clearly seen in Fig.~\ref{fig14} where the lines depicting $L_{1,3}(\boldsymbol{B})$ and $[L_{1,3}(\boldsymbol{B})+L_{3,1}(\boldsymbol{B})]/2$ are the same.  Here, the validity of the relations~\eqref{Casimir} and~\eqref{FDT} for $\alpha = 1$ and $\beta = 3$ is thus demonstrated by the coincidence of the symbols with the solid line.

We then illustrate in Fig.~\ref{fig15} the relation~\eqref{C-M-odd-bis} between the third cumulants $C$ and the second response coefficients $M$ in the case of 4 terminals, respectively for $(\alpha , \beta , \gamma) =(1,1,1)$, $(\alpha , \beta , \gamma)=(1,1,2)$, $(\alpha , \beta , \gamma)=(1,1,3)$ and $(\alpha , \beta , \gamma)=(1,2,3)$. The solid lines show the left-hand sides of Eq.~\eqref{C-M-odd-bis} for a magnetic field $\boldsymbol{B}$. The corresponding quantities are then depicted, for a reversed magnetic field $-\boldsymbol{B}$, by the filled circles, up triangles, down triangles, and diamonds, respectively. The two quantities are here again seen to be the same.

Figure~\ref{fig16} first shows, with the straight horizontal line, that the cumulant $C_{111}$ vanishes identically. The other solid line then shows the cumulant $C_{112}$. These two cumulants are alternatively given by the expressions~\eqref{C0-D1}, \eqref{M-D1-C0}, or~\eqref{C-M-odd} for $(\alpha = \beta =\gamma = 1)$ or $(\alpha = \beta = 1,\gamma = 2)$. The validity of each of these relations, whose right-hand sides are depicted by pluses, open squares and crosses, respectively, is here again fully confirmed. Again, the third cumulants are vanishing at equilibrium: $C_{111}=C_{222}=C_{333}=0$.

We now analyze in Fig.~\ref{fig17} the relation~\eqref{D2-C1} for four different sets of values of the indices $\alpha , \beta , \gamma$, and $\delta$. The solid lines show the left-hand side of the relation~\eqref{D2-C1} for $(\alpha , \beta , \gamma , \delta)=(1,1,1,1)$, $(\alpha , \beta , \gamma , \delta)=(1,1,1,2)$, $(\alpha , \beta , \gamma , \delta)=(1,1,2,2)$, and $(\alpha , \beta , \gamma , \delta)=(1,1,2,3)$. The corresponding right-hand sides are depicted by squares, circles, diamonds, and down triangles, respectively. Thus, the validity of the relation~\eqref{D2-C1} is here also demonstrated.

We conclude in Fig.~\ref{fig18} by considering the relations~\eqref{B0-C1}, \eqref{N-N}, and~\eqref{N-D2-C1-B0}, with the same sets of values of $\alpha , \beta , \gamma$ and $\delta$ as in Fig.~\ref{fig17} above. We recall that these relations are alternative expressions of the fourth cumulants $B$ in terms of the quantities $\partial C / \partial A$, $N$, and/or $\partial^2 D / \partial A^2$. The solid lines show the cumulants $B_{1111}$, $B_{1112}$, $B_{1122}$, and $B_{1123}$. The right-hand sides of the expressions~\eqref{B0-C1}, \eqref{N-N}, and~\eqref{N-D2-C1-B0} of these cumulants are then depicted by open squares, crosses and pluses, respectively. Here again, our analysis confirms the validity of the relations~\eqref{B0-C1}, \eqref{N-N}, and~\eqref{N-D2-C1-B0}.


\section{Conclusion}
\label{Conclusion}

In this paper we analyzed the implications of microreversibility on quantum transport in multi-terminal circuits. The latter consist in $r \geqslant 2$ reservoirs of electrons, coupled through a central region that acts as a scatterer for the electrons. We assume the presence of a nonzero external magnetic field $\boldsymbol{B}$ within the scattering region. Differences of temperatures and chemical potentials in the reservoirs make such a circuit operate out of equilibrium.

The nonequilibrium state is characterized by the occurrence of mean currents of energy and electrons, which are random variables at the microscale. The statistical properties of the currents in the long-time limit are fully specified by the generating function $Q(\boldsymbol{\lambda},\boldsymbol{A};\boldsymbol{B})$ of the statistical cumulants. The latter is a function of the counting parameters $\boldsymbol{\lambda}$ and the affinities $\boldsymbol{A}$ that drive the currents, as well as of the magnetic field $\boldsymbol{B}$.

An interesting feature of the cumulant generating function in multi-terminal circuits is that it can be expressed in terms of the scattering matrix, which describes the scattering processes at the basis of transport. For the kind of systems considered throughout this work the scattering matrix is a $r \times r$ unitary matrix, which is symmetric under the time-reversal transformation. As a consequence, the function $Q$ obeys a particular symmetry relation known as a multivariate fluctuation relation. The latter generates a hierarchy of time-reversal symmetry relations satisfied by the cumulants and their responses to the affinities at arbitrary orders in the deviations from equilibrium.

We illustrated such relations for the first, second, third and fourth cumulants (and their responses) in the particular case of a circuit consisting in $r$ terminals connected through an Aharonov-Bohm ring. A noteworthy advantage of considering such a circuit is that the scattering matrix can be analytically derived. We obtained explicit expressions of the latter in the cases of a ring connected to two, three and four terminals. The knowledge of the scattering matrix then allowed us to construct the cumulants by means of numerical evaluations of integrals. We performed a detailed numerical analysis in order to test the relations between cumulants that are inferred from the fluctuation relation. Our numerical calculations thus fully confirm the validity of such time-reversal relations in the case of an Aharonov-Bohm ring.

Fluctuation relations constitute an important tool for nonequilibrium statistical mechanics, as they are one of the rare exact results that hold arbitrarily far from equilibrium. A deeper understanding of such relations is thus both of fundamental and practical interest. On the fundamental level, we think that our analysis contributes to strengthen the validity of fluctuation relations. Indeed, it has been firmly confirmed on the particular case of Aharonov-Bohm rings. The latter are in addition relevant on the practical level, as they belong to this class of systems that are experimentally realizable in condensed-matter physics. In this perspective, we believe that our work could be of interest for experimental investigations of fluctuation relations based on Aharonov-Bohm rings.


\section*{Acknowledgments}

This research is financially supported by the Universit\'e Libre de Bruxelles (ULB) and the Fonds de la Recherche Scientifique~-~FNRS under the Grant PDR~T.0094.16 for the project ``SYMSTATPHYS".


\appendix

\section{Numerical methods}\label{numerics}

At equilibrium, the chemical potential is common to all the reservoirs and it is fixed to the value $\mu=1$ in our numerical calculations, which are then performed for different values of the temperature.
The cumulants are obtained by integrating~\eqref{Q1}-\eqref{Q3} over the energy $\varepsilon=p^2/(2m)$ with the mass $m=1$ and Planck's constant $\hbar=1$ in de Broglie's formula $p=\hbar k$.  In the units where $\mu=1$, $m=1$, and $\hbar=1$, the perimeter length of the Aharonov-Bohm ring is taken to be $L=2\pi R=8$. Numerical integration is performed by steps $\Delta p = 8/N_p$ with $N_p=20000$, which has been checked to be large enough for effective convergence.  The response coefficients are obtained by differentiating numerically the cumulants with respect to the affinities using the standard formulas \cite{AbrSteg}:
\bea
&&\partial_xf \simeq \frac{1}{2h} \, (f_1-f_{-1}) \, , \\
&&\partial_x^2f \simeq \frac{1}{h^2} \, (f_1-2 f_0+f_{-1}) \, , \\
&&\partial_x^3f \simeq \frac{1}{2h^3} \, (f_2-2f_1+2f_{-1}-f_{-2}) \, , \\
&&\partial_x\partial_yf \simeq \frac{1}{4h^2} \, (f_{1,1}-f_{1,-1}-f_{-1,1}+f_{-1,-1}) \, , 
\eea
where $f_j\equiv f(x+j h)$, $f_{j,k}=f(x+jh,y+kh)$, and $h=\Delta x= \Delta y$.  For differentiating with respect to the affinities, the step $h=\Delta A=\Delta\mu/(k_{\rm B}T)$ is taken for every affinity $A_{\alpha}$ considered. The step value $\Delta\mu=10^{-6}$ is used to obtain the response coefficients $L_{\alpha,\beta}$ and $M_{\alpha,\beta\gamma}$, as well as for the first derivatives of the cumulants $D_{\alpha\beta}$ and $C_{\alpha\beta\gamma}$ with respect to the affinities. The value $\Delta\mu=10^{-3}$ is used for the response coefficients $N_{\alpha,\beta\gamma\delta}$ and for the second derivatives of the cumulants $D_{\alpha\beta}$ with respect to the affinities.  All the quantities are computed for values of the dimensionless magnetic flux separated by $\Delta\phi=10^{-2}$.

%


\end{document}